\documentclass[12pt,preprint]{aastex}
\usepackage{amsmath}
\shorttitle{Photosphere emission in X-Ray Flares of GRBs}
\begin{document}

\title{Photosphere emission in the X-Ray Flares of {\em Swift} Gamma-Ray Bursts and Implications for the Fireball Properties}
\author{Fang-Kun Peng\altaffilmark{1,2,3}, En-Wei Liang\altaffilmark{1,3,4}, Xiang-Yu Wang\altaffilmark{2}, Shu-Jin Hou\altaffilmark{5,6}, Shao-Qiang Xi\altaffilmark{1}, Rui-Jing Lu\altaffilmark{1}, Jin Zhang\altaffilmark{4}, Bing Zhang\altaffilmark{7,1}
}
\altaffiltext{1}{Department of Physics and GXU-NAOC Center for Astrophysics
and Space Sciences, Guangxi University, Nanning 530004,
China; lew@gxu.edu.cn}
\altaffiltext{2}{School of Astronomy and Space Science, Nanjing University, Nanjing 210093, China; xywang@nju.edu.cn}
\altaffiltext{3}{Key Laboratory for the Structure and Evolution of Celestial Objects, CAS, Kuming 650011}
\altaffiltext{4}{National Astronomical Observatories, Chinese Academy of Sciences, Beijing, 100012, China}
\altaffiltext{5}{Institute of Physics and Electronic Engineering, Nanyang Normal College, Nanyang, 473061, China}
\altaffiltext{6}{Purple Mountain Observatory, Chinese Academy of Sciences, Nanjing, 210008, China}
\altaffiltext{7}{Department of Physics and Astronomy, University of
Nevada, Las Vegas, NV 89154; zhang@physics.unlv.edu}
\begin{abstract}
X-ray flares of gamma-ray bursts (GRBs) are usually observed in
the soft X-ray range and the spectral coverage is limited. In this
paper, we present an analysis of 32 GRB X-ray flares that are
simultaneously observed by both BAT and XRT on board the {\em Swift} mission, so a joint spectral analysis with a wider spectral
coverage is possible. Our results show that the joint spectra of
19 flares are fitted with the absorbed single power-law or the
Band function models. More interestingly, the
joint spectra of the other 13 X-ray flares are fitted with the
absorbed single power-law model plus a black body (BB)
component. Phenomenally, the observed spectra of these 13 flares
are analogous to several GRBs with a thermal component, but only with a much
lower temperature of $kT=1\sim 3$ keV. Assuming that the thermal
emission is the photosphere emission of the GRB fireball, we
derive the fireball properties of the 13 flares that have redshift
measurements, such as the bulk Lorentz factor  $\Gamma_{\rm ph}$
of the outflow. The derived $\Gamma_{\rm ph}$ range from $50$ to $150$ and a relation of $\Gamma_{\rm ph}$ to the thermal emission luminosity is found. It is consistent with the $\Gamma_0-L_{\rm iso}$ relation that are derived for the prompt gamma-ray
emission. We discuss the physical implications of these results
within the content of jet composition and radiation mechanism
of GRBs and X-ray flares.
\end{abstract}

\keywords{gamma-ray: bursts--X-ray: flare--radiation: thermal and non-thermal }

\section{Introduction}
\label{sec:intro}
The radiation physics of prompt gamma-ray emission is still a mystery. It is related to the unknown composition and radiation mechanism of the GRB jets (e.g. Zhang 2014).
The standard fireball-shock model predicts a GRB prompt emission spectrum as superposition of a quasi-thermal photosphere emission component and a
non-thermal component from internal shocks (e.g., M\'esz\'aros \& Rees 2000; Zhang \& M\'{e}sz\'{a}ros 2002; Daigne \& Mochkovitch 2002; Pe'er et al. 2006; Toma et al. 2011; Pe'er et al. 2012). GRB observations with CGRO/BATSE in the 25-2000 keV band revealed that a typical GRB spectrum in the BATSE band is empirically fitted with a smooth broken power-law function, the so-called ``Band function" (Band et al. 1993; Preece et al. 2000). Possible superposition of a thermal component on the non-thermal spectrum was claimed in some BATSE GRBs (e.g., Ghirlanda et al. 2003; Ryde 2004, 2005; Bosnjak et al. 2006; Ryde \& Pe'er 2009). With the gamma-ray burst monitor (GBM) and large area telescope (LAT) on board the {\em Fermi} mission, GRBs are now observed in a broad spectral band from several keV to 300 GeV. Most {\em Fermi} GRB spectra are still well fitted with the Band function (Abdo et al. 2009; Zhang et al. 2011; Lu et al. 2012). This has led to the suggestion that the composition of GRB jets is magnetically dominated (e.g. Zhang \& Pe'er 2009). However, evidence for a black body (BB) component is also found. The most prominent case is GRB 090902B, whose time-integrated broad band spectrum can be decomposed into a dominating multi-color BB comment and an extra power-law component in the GBM and LAT band (Ryde et al. 2010). For small enough time bins, the time-resolved spectrum can be decomposed into a BB component plus a power law (Zhang et al. 2011). Another interesting case with detection of a BB component is GRB 081221. The spectrum of GRB 081221 shows a bimodal feature, which is well fit with the Band function plus a BB component with a typical $kT\sim 20$ keV (Hou et al. 2014, in preparation). A weak BB component may also contribute to the total flux with a fraction of less than $10 \%$ in GRBs 110721A (Axelsson et al. 2012), 100724B (Guiriec et al. 2011) and GRB 120323A (Guiriec et al. 2013). These observations indicate that the intrinsic GRB spectra of some GRBs are composed of a thermal (or quasi-thermal) component and a non-thermal component (broad Band function or exponential cutoff power-law function), as expected in the standard fireball model.

The \emph{Swift} mission plays a critical role in revealing the nature of the gamma-ray bursts (GRBs) with its rapidly slewing capacity and multi-wavelength observations (Gehrels et al. 2004). Bright X-ray flares have been observed with the X-ray telescopes (XRT, Burrows et al. 2005b) on board {\em Swift} from tens to $10^5$ seconds post GRB triggers with the burst alert telescope (BAT; Barthelmy et al. 2005) for half of GRBs (Burrows et al. 2005a; Falcone et al. 2006; Chincarini et al. 2007, 2010). The majority of flares happened at $t<1000$ seconds, and some flares may occur at $\sim 10^5$ seconds post the triggers. Both spectral and temporary properties of X-ray flares have been extensively studied (e.g. O'Brien et al. 2006; Liang et al. 2006; Falcone et al. 2007; Chincarini et al. 2010; Margutti et al. 2011; Qin et al. 2013; Wang \& Dai 2013; Hu et al. 2014). It is generally believed that X-ray flares are of an internal origin and may signal the restart of the GRB  central engine post the prompt gamma-ray phase (e.g., Burrows et al. 2005a; Zhang et al. 2006; Liang et al. 2006; Dai et al. 2006; see review by Zhang 2007). The spectra of most X-ray flares in the XRT band is adequate to be fitted by an absorbed single power-law model, although the GRB Band function or the single power-law plus a BB component may improve the fits for some flares (Falcone et al. 2007; Page et al. 2011). Chincarini et al. (2010) confirmed that the X-ray  flares are tightly linked to the prompt emission by analyzing the width evolution with energy, the ratio between rising and decaying time scales, as well as the spectral energy distribution of X-ray flares. Margutti et al. (2010) found that the peak energies ($E_{\rm p}$) of the early flares observed in GRBs 060904B and 060418 are marginally consistent with the $E_{\rm p}-L_{\rm iso}$ relation derived from prompt gamma-rays (Yonetoku et al. 2004; Liang et al. 2004), but deviate from the $E_{\rm p}-E_{\gamma, \rm iso}$ relation for typical GRBs (Amati et al. 2002).

This paper is dedicated to search for a BB component in the joint XRT+BAT band (0.3-150 keV) for the observed X-ray flares with the {\em Swift} mission in a sample as described in \S 2. We present a joint spectral analysis of the X-ray flares that were simultaneously observed with BAT and XRT and find a thermal emission component embedded the joint spectra of a fraction of flares (\S 3). By applying the standard fireball photosphere theory (\S 4.1), we constrain the fireball properties of these flares, and found tight $\Gamma_{\rm ph}-L_{\rm BB}$, $E_{\rm p}-L_{\rm BB}$ correlations for the thermal component of these flares (\S 4.2). We draw conclusions in \S5, and discuss the implications for understanding the physics of X-ray flares.
The concordance cosmology with parameters $ H_{0}=71 {\rm km s}^{-1} {\rm
Mpc}^{-1}$,$\Omega_{\rm M}=0.3$ and $\Omega_{\Lambda}=0.7$ is adopted.

\section{Sample Selection and Data Reduction}
\begin{figure}
\includegraphics[angle=0,scale=0.3]{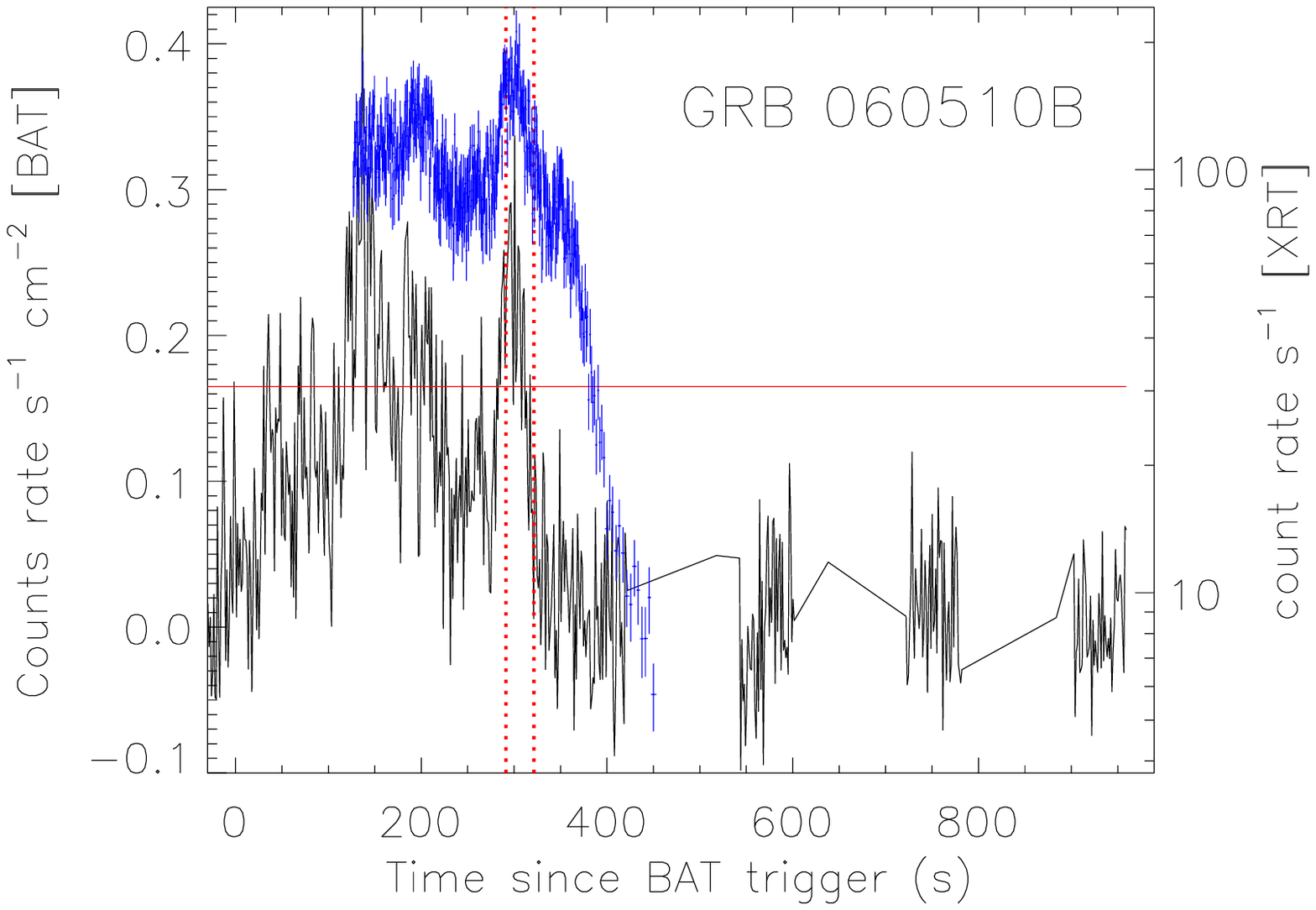}%
\includegraphics[angle=0,scale=0.3]{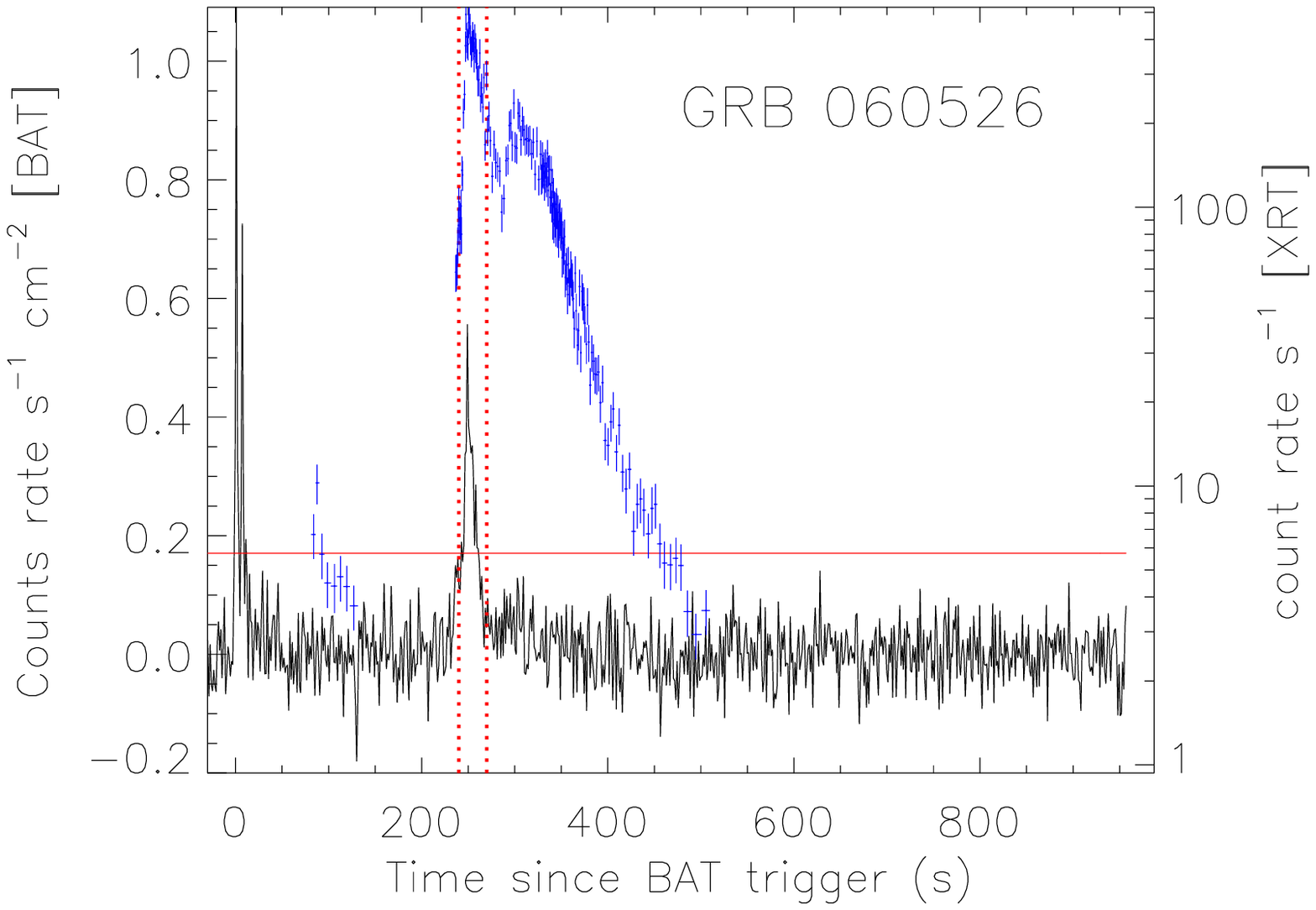}%
\includegraphics[angle=0,scale=0.3]{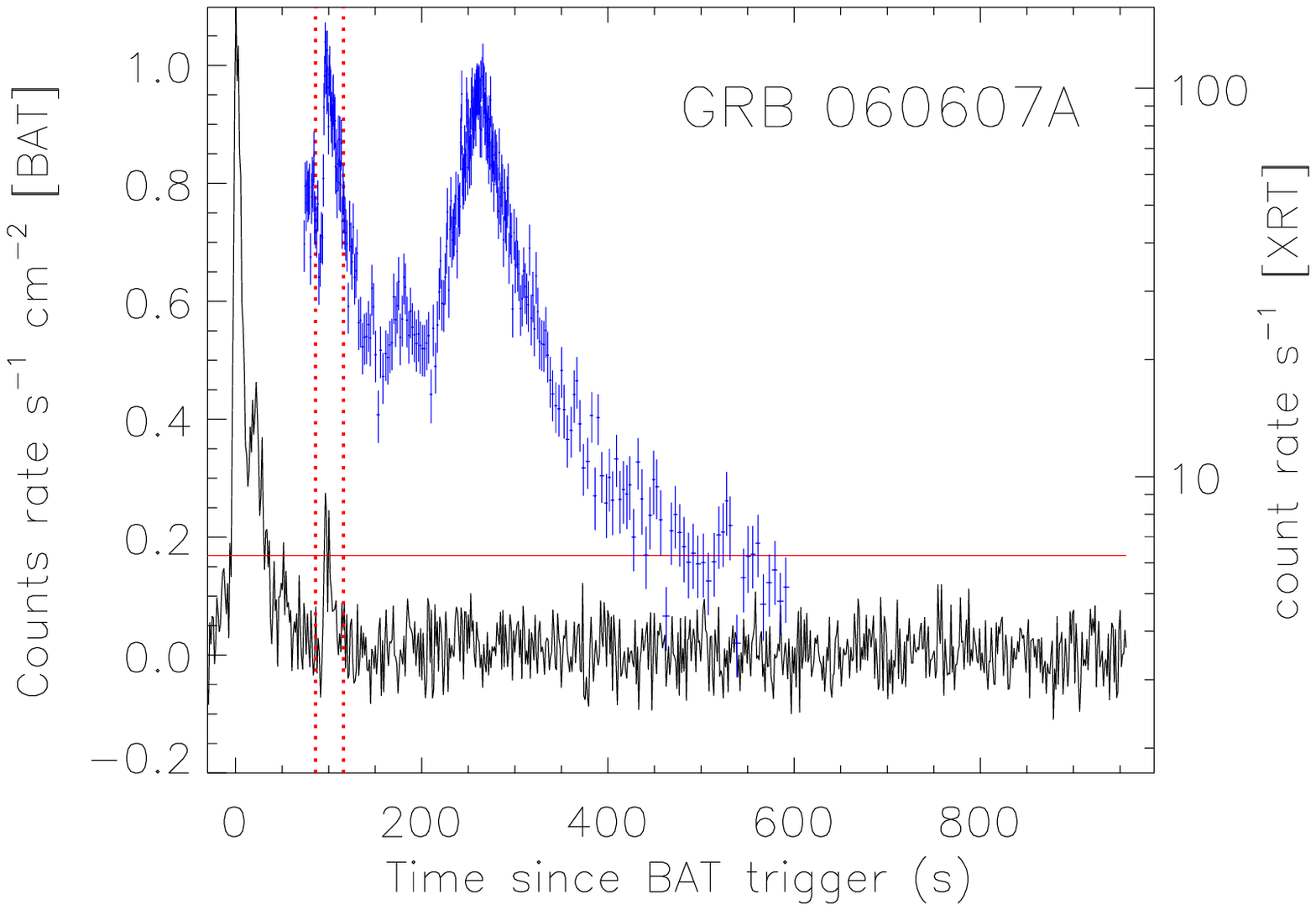}\\
\includegraphics[angle=0,scale=0.3]{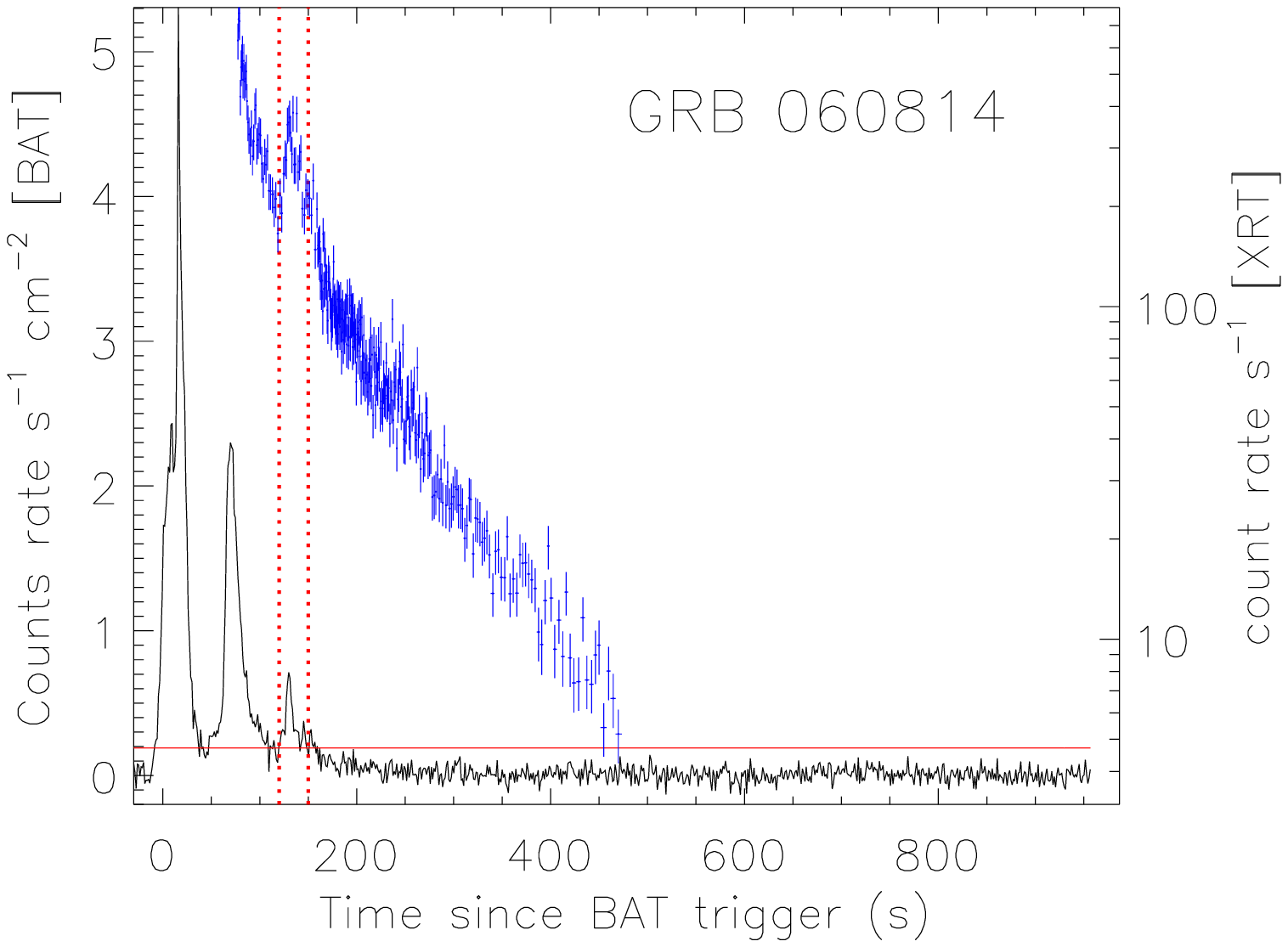}%
\includegraphics[angle=0,scale=0.3]{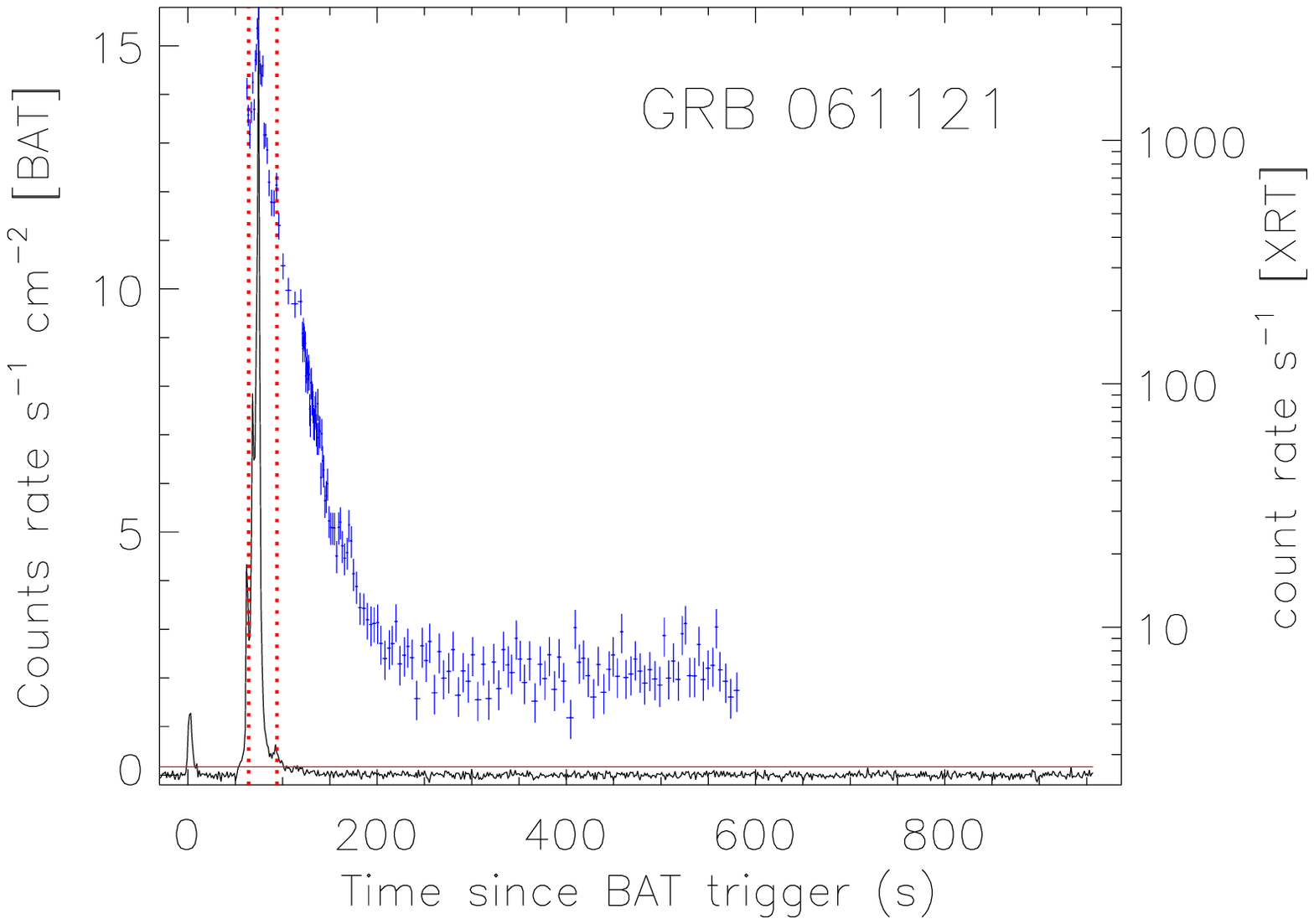}%
\includegraphics[angle=0,scale=0.3]{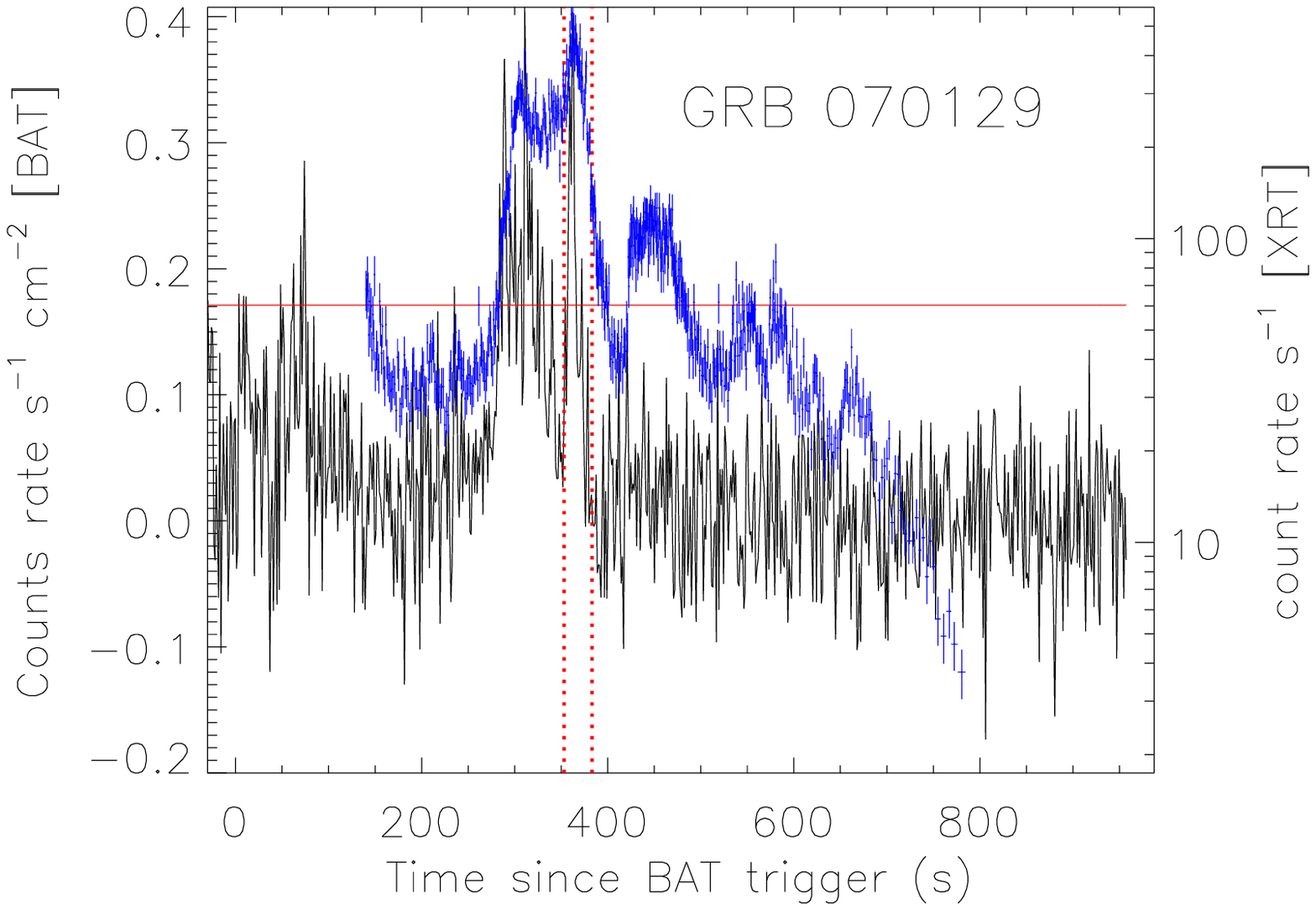}\\
\includegraphics[angle=0,scale=0.3]{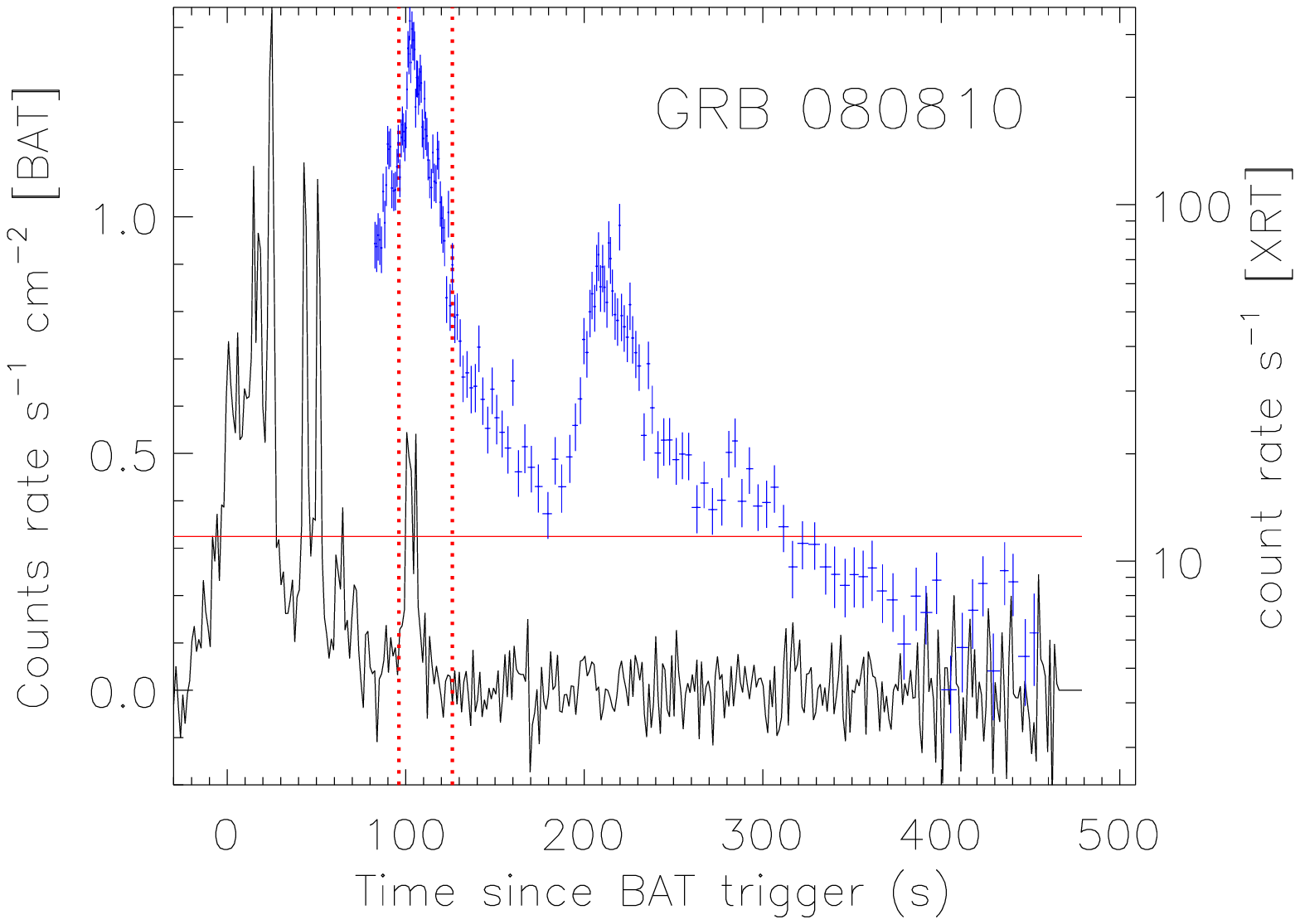}%
\includegraphics[angle=0,scale=0.3]{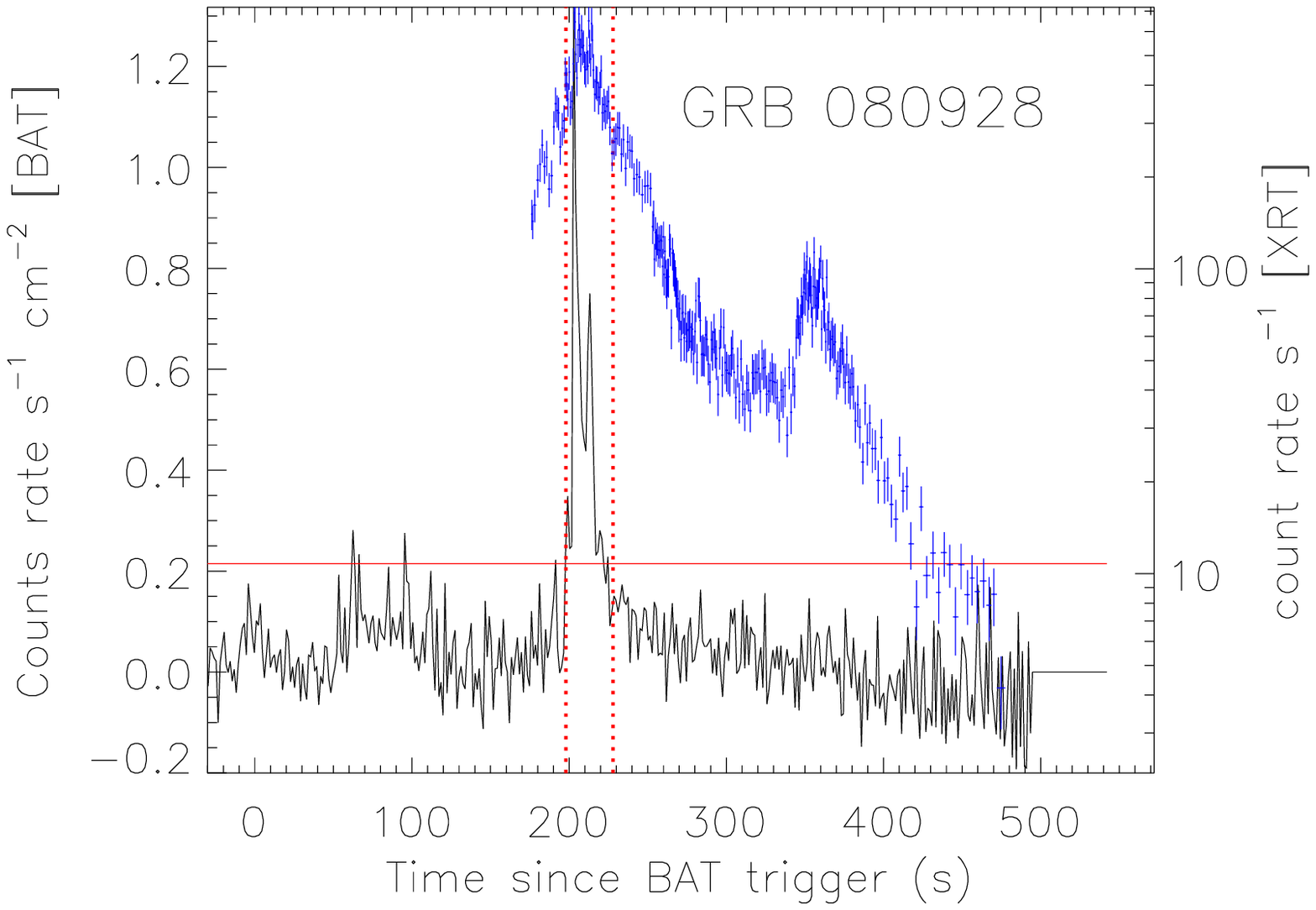}%
\includegraphics[angle=0,scale=0.3]{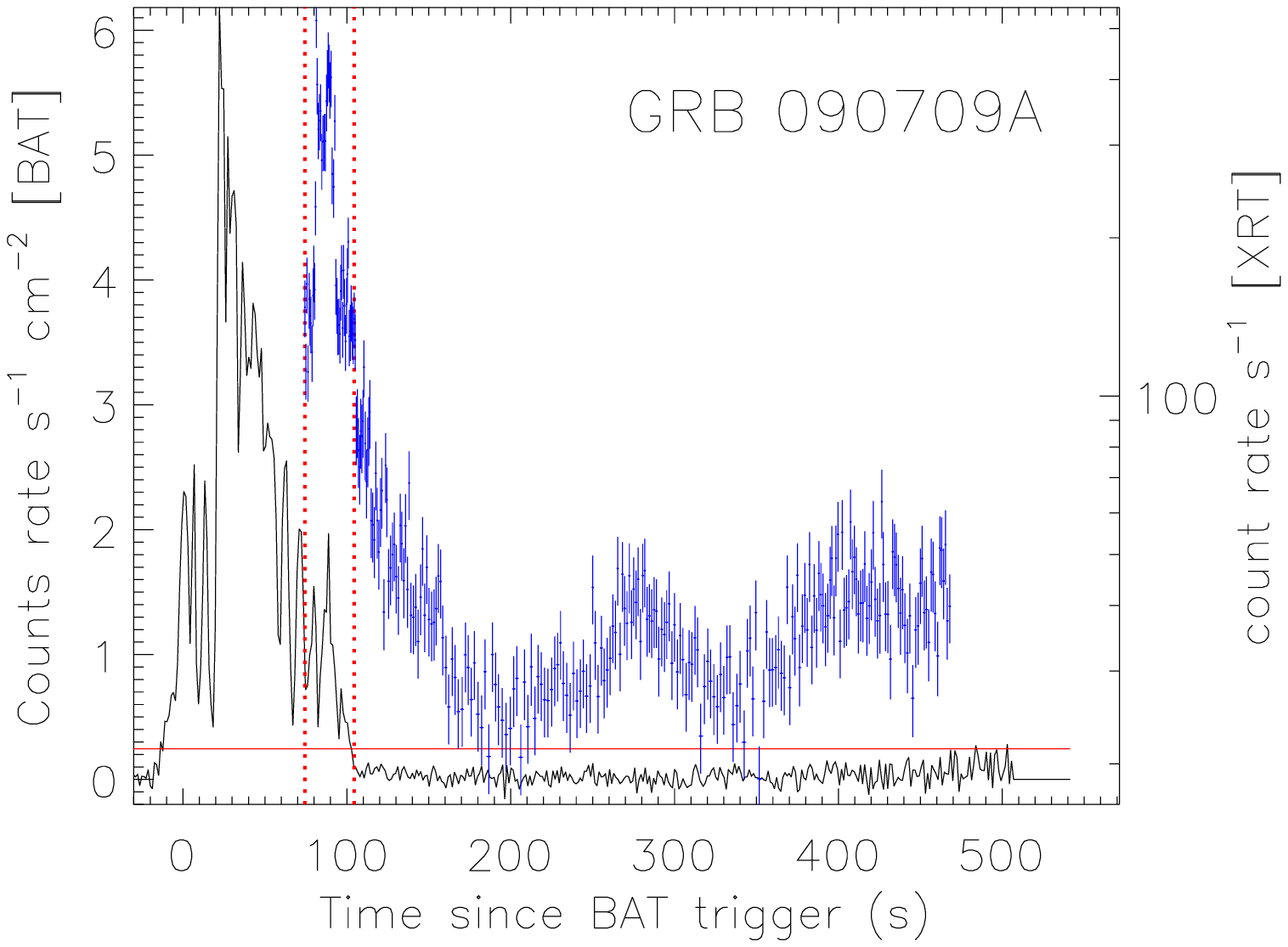}\\
\includegraphics[angle=0,scale=0.3]{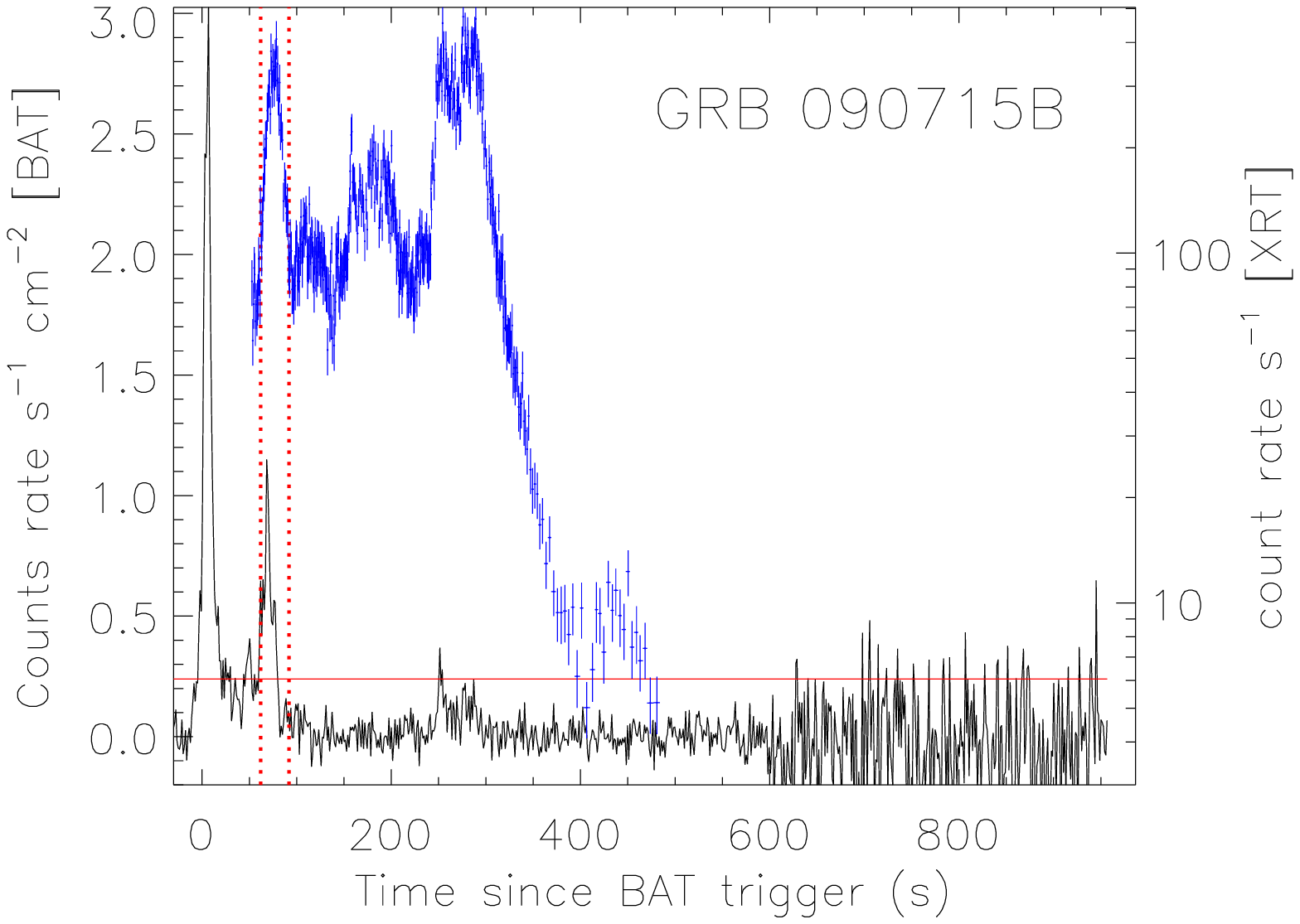}%
\includegraphics[angle=0,scale=0.3]{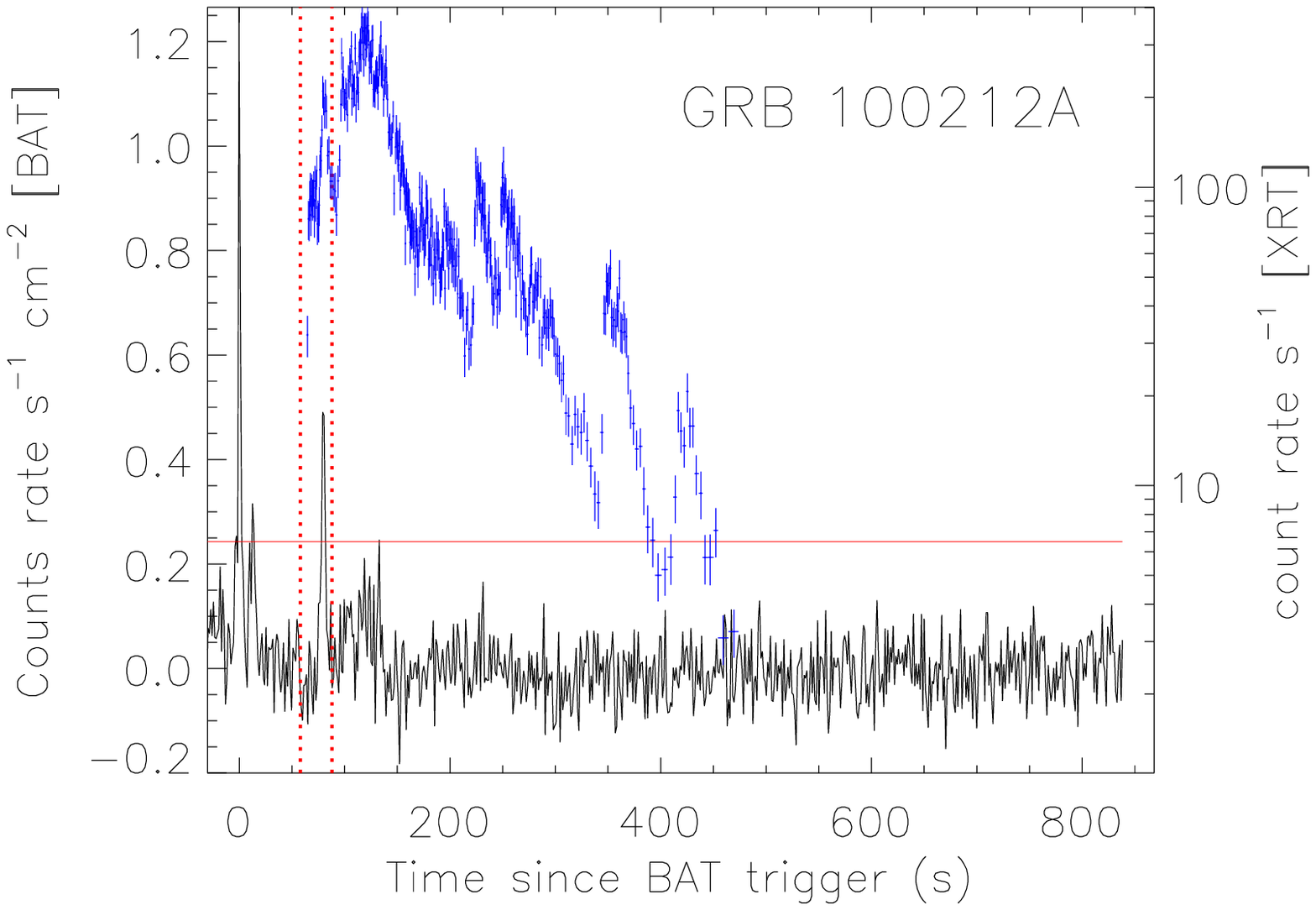}%
\includegraphics[angle=0,scale=0.3]{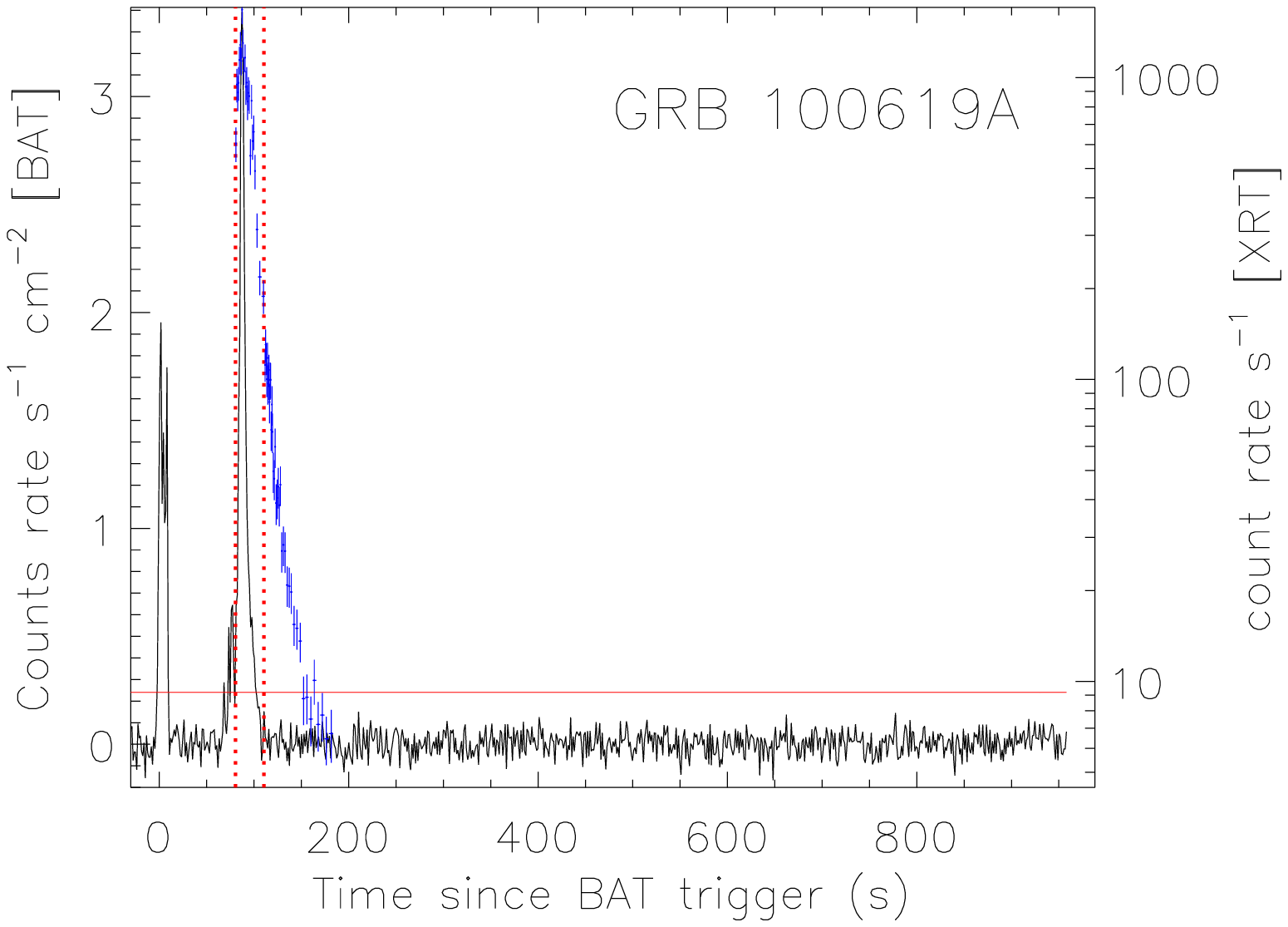}\\
\includegraphics[angle=0,scale=0.3]{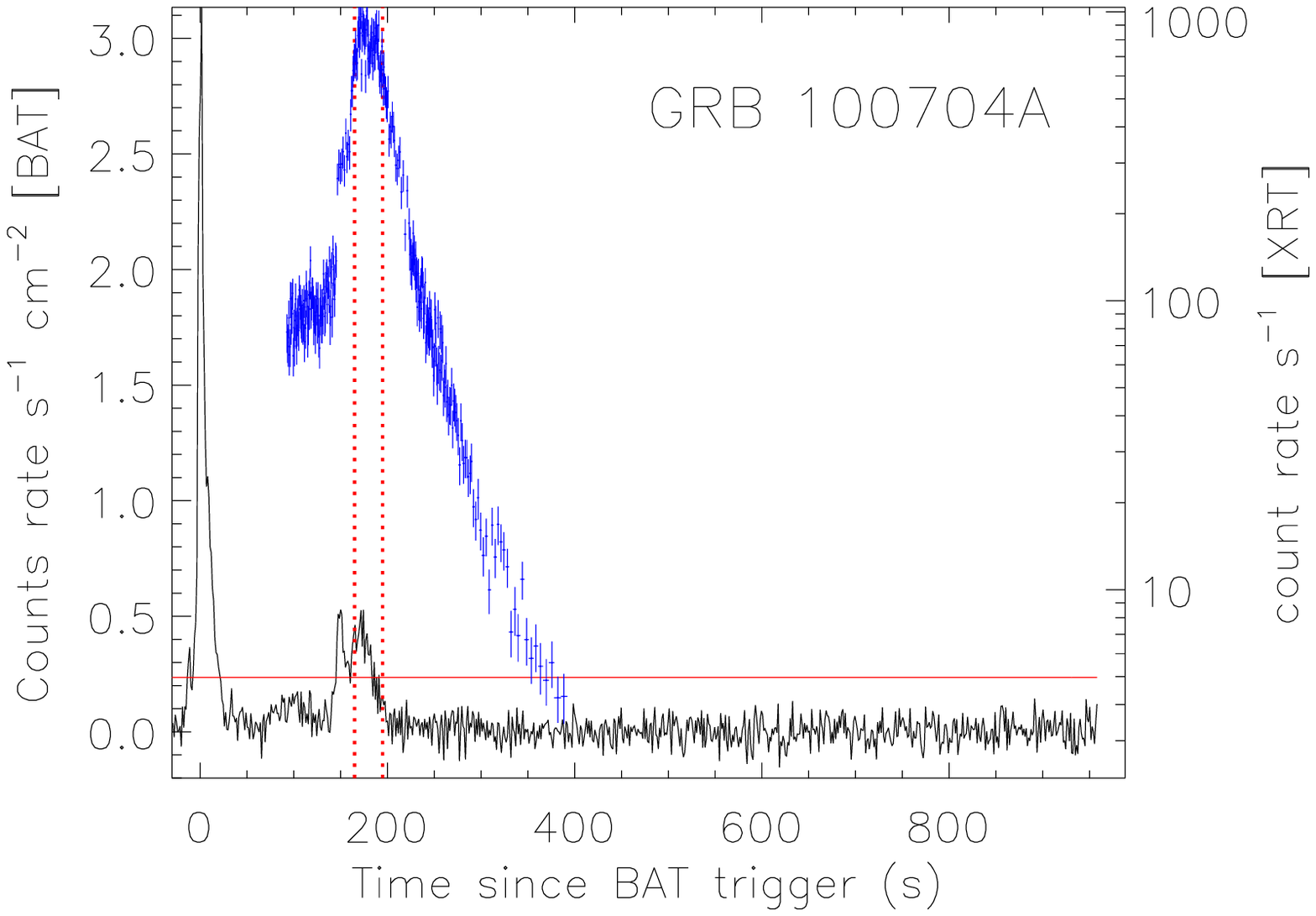}%
\includegraphics[angle=0,scale=0.3]{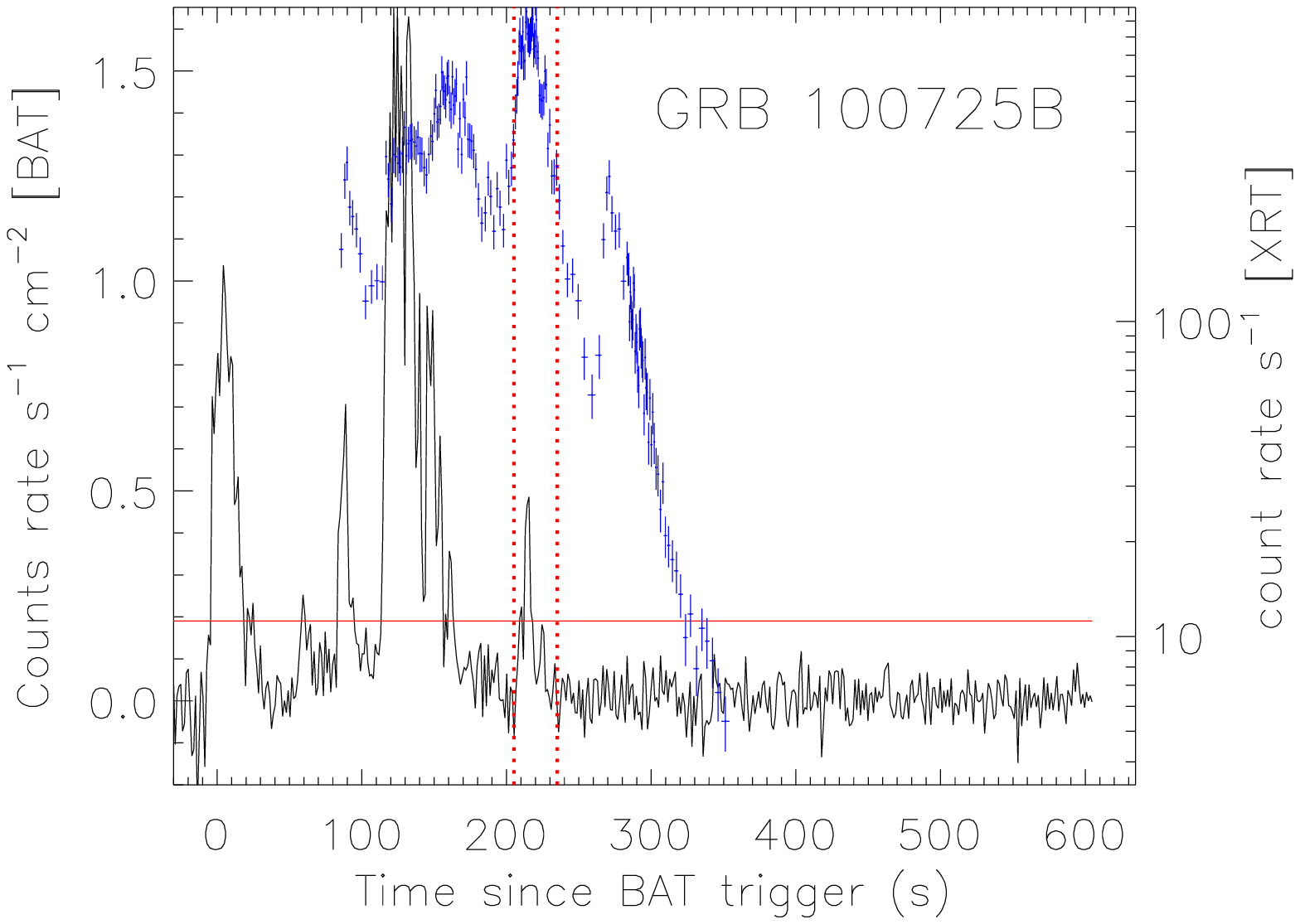}%
\includegraphics[angle=0,scale=0.3]{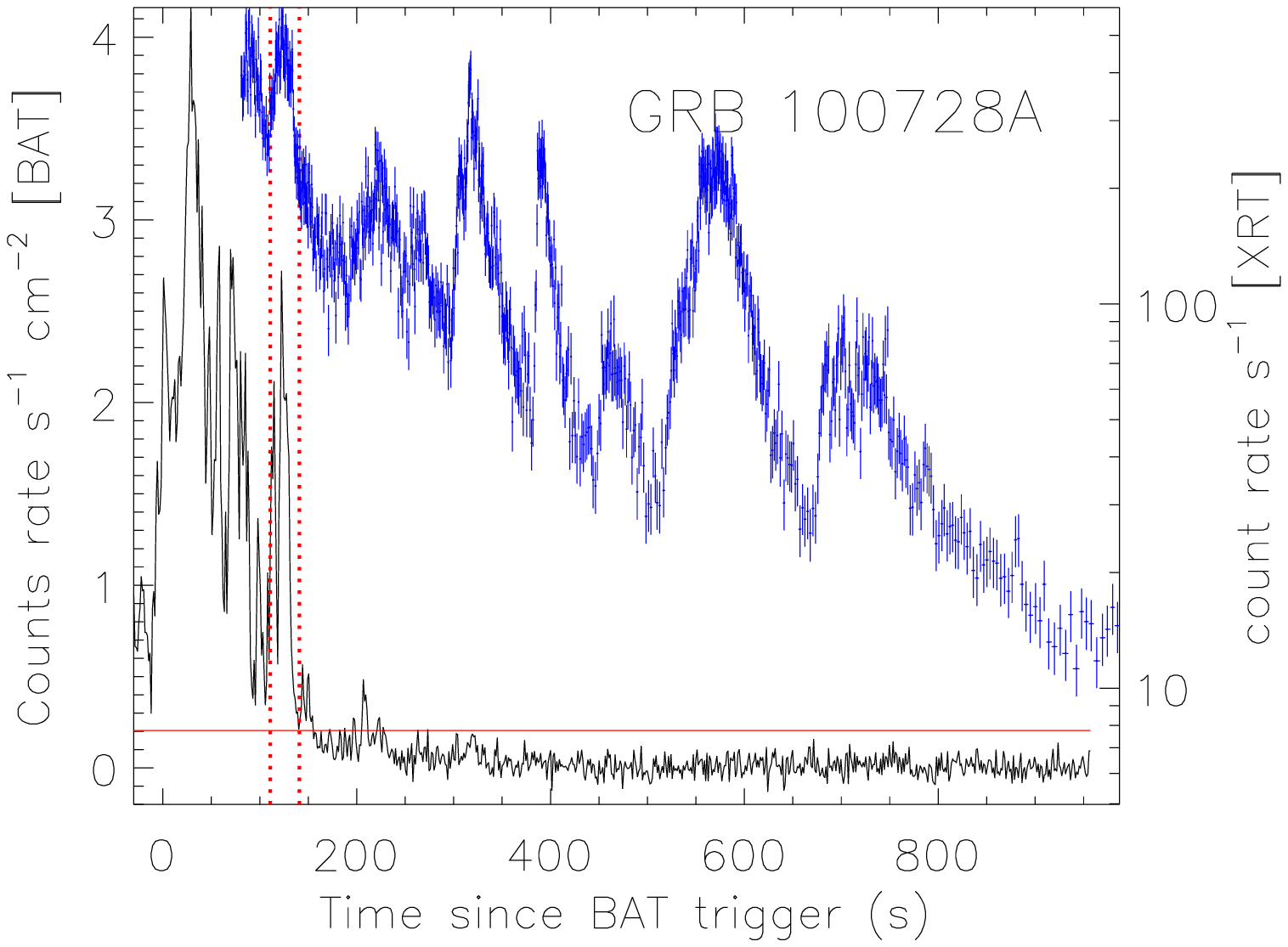}\\
\hfill
\caption{Joint lightcurves observed with XRT (blue crosses) and BAT (black connected lines) for the GRBs in our sample. The selected intervals for our spectral analysis are marked with red vertical dotted lines. The BAT count rate and the XRT count rate in each panel are shown in linear (left vertical axis) and log (right vertical axis) scales, respectively. The horizonal dashed lines mark the 4 $\sigma$ level of the BAT background.}
\label{LCs}
\end{figure}

With the promptly slewing capacity of the Swift mission, the prompt emission of some GRBs was simultaneously observed with BAT and XRT. We search for the flares that are also bright in the BAT band to make joint spectral fits. We take the XRT lightcurves of the GRBs observed with Swift from http://www.swift.ac.uk/xrt\_spectra/ (Evans et al. 2009). The XRT lightcurves are binned dynamically with a minimum signal to noise ratio of 3. The XRT lightcurves are usually composed of an underlying component with multiple power-law-decaying segments and some flares (Zhang et al. 2006; Nousek et al. 2006). We select only those flares that do not significantly overlap with adjacent flares. We empirically fit the lightcurves around the selected flares with a model of a broken power-law plus a single power-law to obtain the profiles of the flares (Margutti et al. 2010, 2011; Chincarini et al. 2010.), similar to what was usually done in decomposing the prompt gamma-ray pulses (e.g., Norris et al. 1996). We extract the background-subtracted BAT lightcurves with a bin size of 1.024s, and use the Beyasian black method to analyze the profile of the BAT lightcurves. For the details of our analysis on the BAT lightcurves, please refer to Hu et al. (2014). We adopt the following criteria to select the flares for our analysis. First, the flares are bright, with $F_{\rm p}/F_{\rm u}>5$, where $F_{\rm p}$ is the flux at the peak time ($t_{\rm p}$) and $F_{\rm u}$ is the flux of the underlying power-law component at the peak time derived from our empirical fits. Since X-ray flares are asymmetric with a slow decay and relatively fast
rising wing (Chincarini et al. 2010), we normally select a time interval $[t_{\rm p}-10 s,t_{\rm p}+20 s]$ for our analysis, but sometimes the time interval may vary
based on the detailed lightcurve behavior. Second, the BAT lightcurves in the corresponding time intervals have a signal to noise ratio being greater than $4\sigma$, where $\sigma$ is the standard deviation of the background. With these criteria the extracted spectra are dominated by the flares/pulses in the XRT-BAT band. Based on these criteria, we identified a sample of 32 bright flares for our analysis by the end of April, 2014.\footnote{Bright flares in GRB 060124 are excluded since no simultaneous BAT event data are available during the flares for this GRBs (Romano et al. 2006). Since we focus on the joint BAT-XRT data analysis in this paper, we therefore do not include this event in our sample.}. These flares belong to different GRBs, so we finally get a sample of 32 GRBs. Their joint BAT and XRT lightcurves and the selected time intervals are illustrated in Figure \ref{LCs}. Associated peaks in the BAT and XRT bands are found in our selected flares, although the corresponding X-ray flares are usually much broader than the pulses in the BAT band.

We extract the BAT spectrum with the HEAsoft package (ver.6.15). The BAT event data are downloaded from \emph{Swift} archive\footnote{http://heasarc.gsfc.nasa.gov/cgi-bin/W3Browse/swift.pl.}. We use the tool \emph{batbinevt} to extract the spectrum and apply corrections to the spectra with tools \emph{batupdatephakw} and \emph{batphasyserr}. A response matrix then is generated with \emph{batdrmgen}. The XRT spectra are extracted using an interacting tool available from the XRT team\footnote{website at http://www.swift.ac.uk/xrt\_spectra/ (Evans et al. 2009).}. We bin the XRT spectrum at least 20 counts per bin with a tool \emph{grppha}.

\section{Joint Spectral Fits}
The derived BAT and XRT spectra are jointly fitted with the spectral analysis package Xspec (version: 12.8.1). The Galactic absorption is fixed (Kalberla et al. 2005) and the intrinsic absorption is adopted as a free parameter. Three spectral models are considered, i.e.,

(1) the single power-law model (PL model),
\begin{equation}
 { N(E)={ K_{\rm PL}} E^{-\Gamma}};
\end{equation}

(2) the black body radiation model (BB model),
\begin{equation}
{ A(E)=\frac{8.0525 \times K_{\rm BB}  E^2 d E}{(kT)^4 (e^{(E/kT)}-1)}};
\end{equation}
and (3) the Band function,
\begin{equation}
   A(E)=
   \begin{cases}
    K(E/100)^{\alpha}e^{-E/E_0} &\mbox{if $E<(\alpha-\beta)E_0$ }\\
    K[(\alpha-\beta)E_c/100]^{(\alpha-\beta)}(E/100)^{\beta}e^{-(\alpha-\beta)} &\mbox{if $E > (\alpha-\beta)E_0$}
   \end{cases}.
\end{equation}

The strategy and procedure of our spectral fits are described as the following.

\begin{itemize}
\item  Since the observed spectra in the 0.3-2 keV band highly suffer absorbtion by metals (denoted as the corresponding neutral hydrogen column density $N_{\rm H}$) along the line of sight, we first make a preliminary fit to a joint XRT+BAT spectrum in the 2-150 keV band with the PL model in order to judge the preferred models we use.

\item We inspect the deviation between the observed spectrum and the absorbed PL fit to select preferred models. In the case of that the model is adequate to fit the spectrum in the 2-150 keV band, with a reduced $\chi^2<1.1$, the preferred model remains as the absorbed PL model. In the case that both the high energy ($>40$ keV) end and the low energy ($2-5$ keV) end of the spectrum are consistent with the fitting line, but the data show a bump feature in the middle range, the preferred model to improve the fit is the absorbed single power-law plus a BB component (the PL+BB model). In the case that observed spectrum is curved, with both the high energy ($>40$ keV) end and low energy ($\sim 2-5$ keV) end of the spectrum deviate from the single power law fitting line, the preferred model to improve the fit is the absorbed Band function.

\item We make global fits to the observed spectrum in the 0.3-150 keV band with the preferred models and compare the results to the fit with the absorbed PL model. If the parameters of the preferred models are well constrained and the reduced $\chi^2$ is $\sim 1.0$, we adopt the fitting results of the preferred models.
\end{itemize}

\begin{figure}
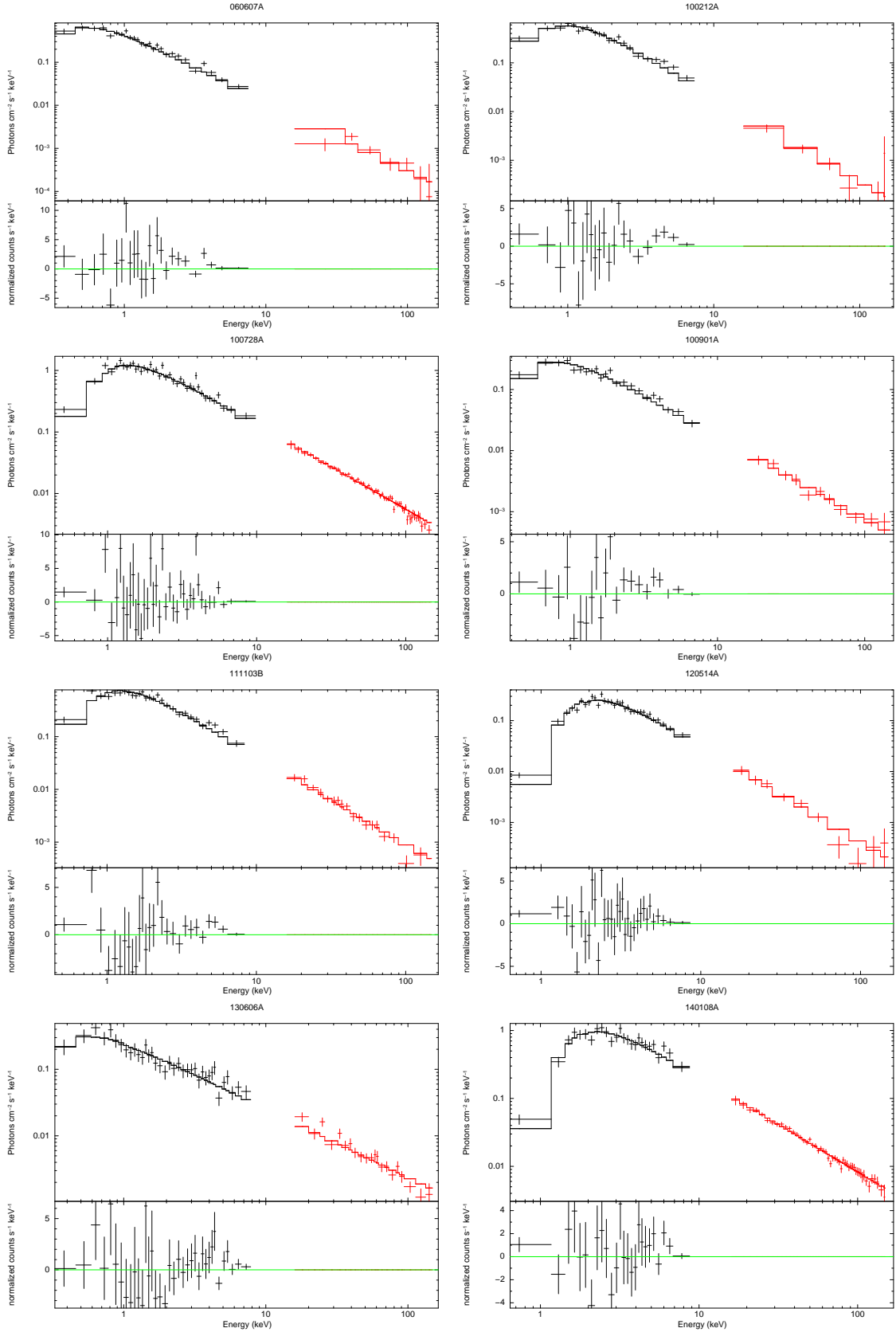

\includegraphics[angle=270,scale=0.3]{060607A.eps}%
\includegraphics[angle=270,scale=0.3]{100212A.eps}\\
\includegraphics[angle=270,scale=0.3]{100728A.eps}%
\includegraphics[angle=270,scale=0.3]{100901A.eps}\\
\includegraphics[angle=270,scale=0.3]{111103B.eps}%
\includegraphics[angle=270,scale=0.3]{120514A.eps}\\
\includegraphics[angle=270,scale=0.3]{130606A.eps}%
\includegraphics[angle=270,scale=0.3]{140108A.eps}\\
\hfill
\caption{Joint BAT and XRT spectra with our single power-law fit for 8 GRBs reported in Table 1.}
\label{PL}
\end{figure}

\begin{figure}
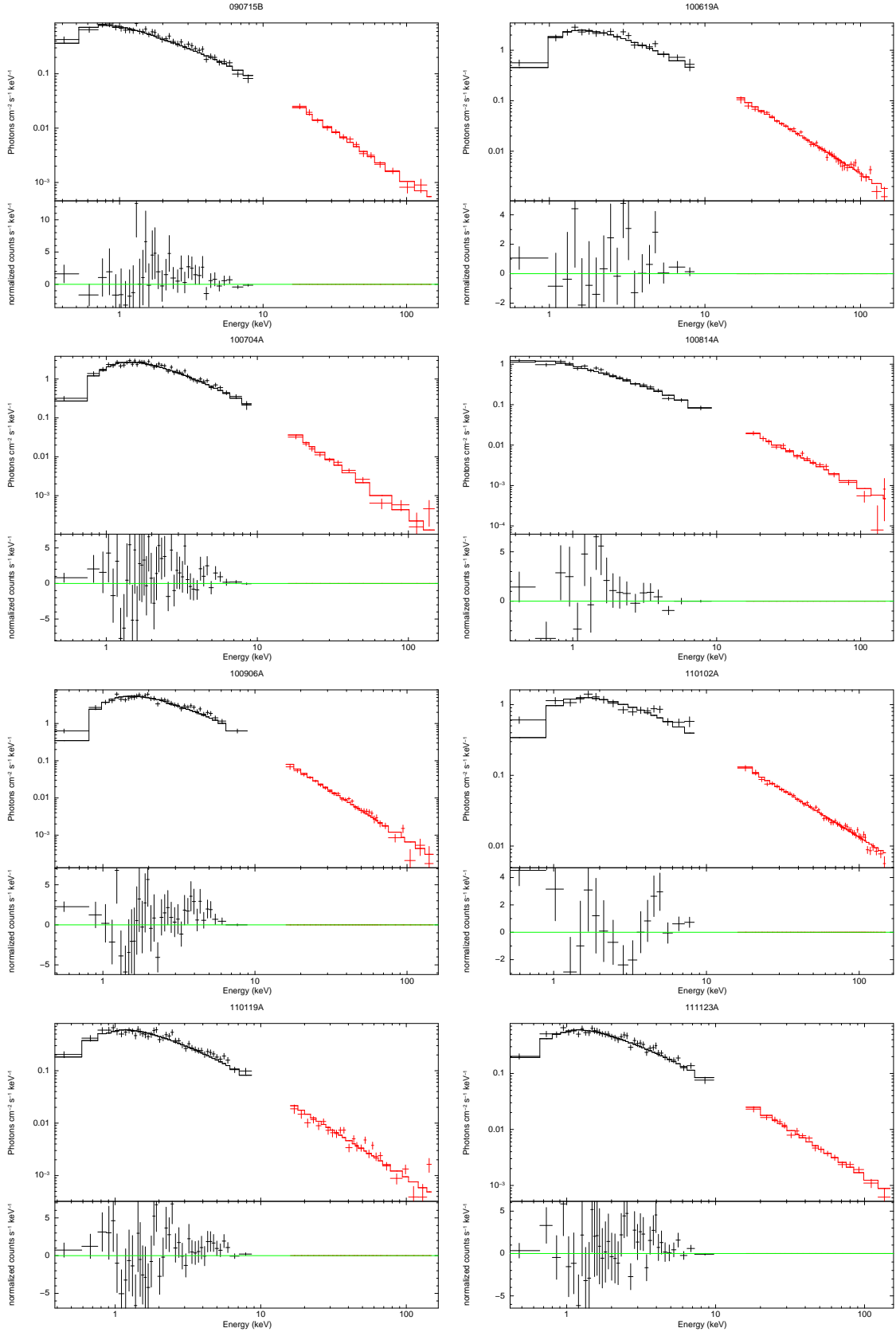

\includegraphics[angle=270,scale=0.3]{090715B.eps}%
\includegraphics[angle=270,scale=0.3]{100619A.eps}\\
\includegraphics[angle=270,scale=0.3]{100704A.eps}%
\includegraphics[angle=270,scale=0.3]{100814A.eps}\\
\includegraphics[angle=270,scale=0.3]{100906A.eps}%
\includegraphics[angle=270,scale=0.3]{110102A.eps}\\
\includegraphics[angle=270,scale=0.3]{110119A.eps}%
\includegraphics[angle=270,scale=0.3]{111123A.eps}\\
\hfill
\caption{Joint BAT and XRT spectra together with the Band function fits for 11 GRBs reported in Table 2.}
\label{Band}
\end{figure}

\begin{figure}
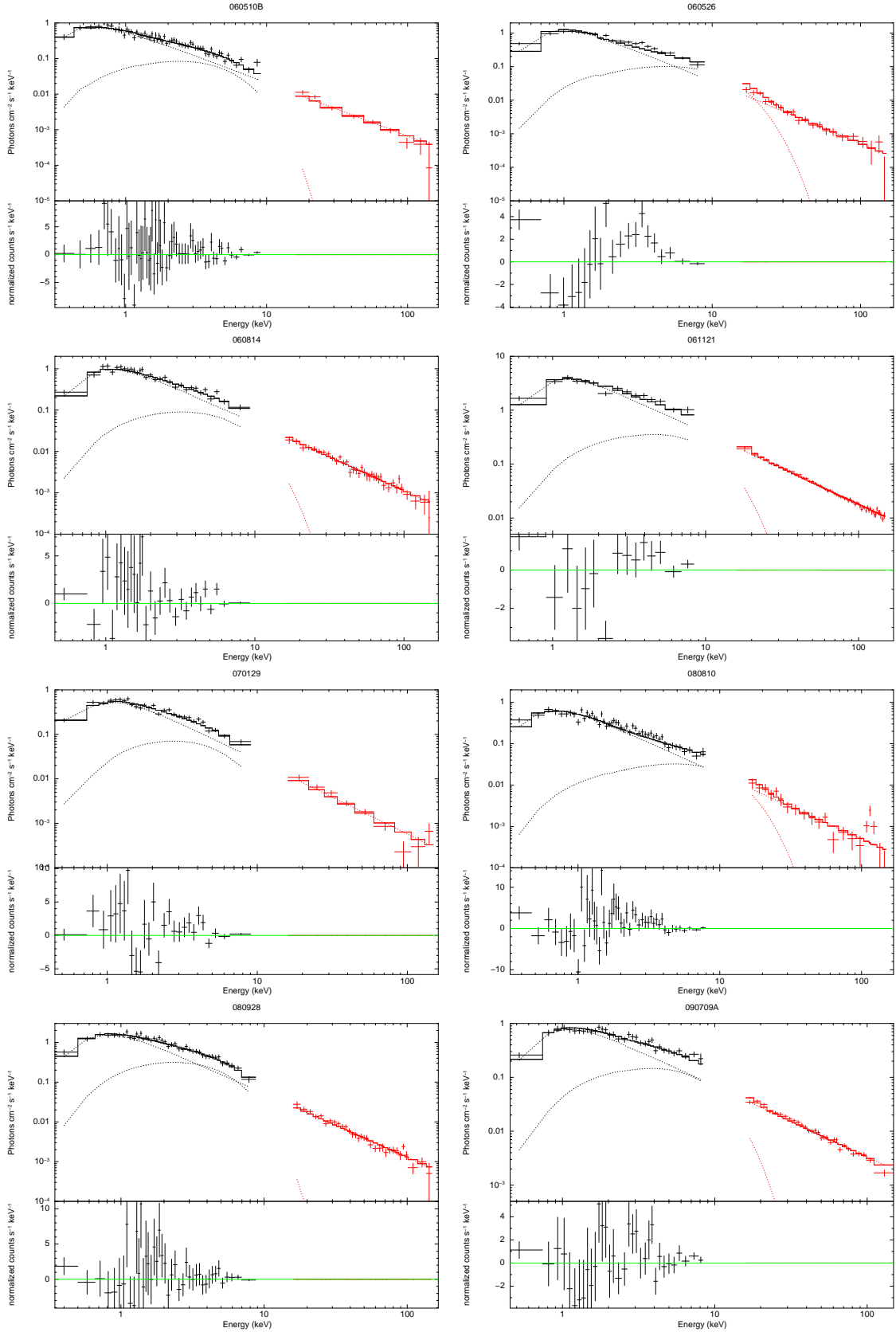

\includegraphics[angle=270,scale=0.3]{060510B.eps}%
\includegraphics[angle=270,scale=0.3]{060526.eps}\\
\includegraphics[angle=270,scale=0.3]{060814.eps}%
\includegraphics[angle=270,scale=0.3]{061121.eps}\\
\includegraphics[angle=270,scale=0.3]{070129.eps}%
\includegraphics[angle=270,scale=0.3]{080810.eps}\\
\includegraphics[angle=270,scale=0.3]{080928.eps}%
\includegraphics[angle=270,scale=0.3]{090709A.eps}\\
\hfill
\caption{Joint BAT and XRT spectra together with the PL+BB model fits for 13 GRBs reported in Table 3.}
\label{PL+BB}
\end{figure}

\begin{figure}
\centering
\includegraphics[angle=0,scale=0.5]{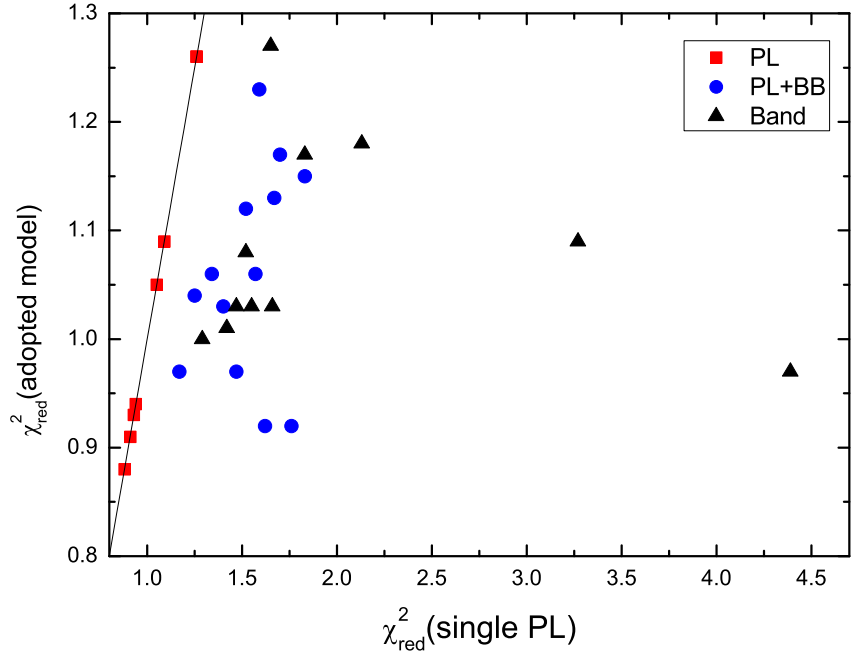}%
\caption{Comparison of the reduced $\chi^2$ of the fits between the adopted model and the single power-law model. A single power-law fit with $\chi^2_r<1.3$ is normally adopted, except when the PL+BL or the Band function models can reduce the $\chi^2_{\rm r}$ down to  $<1.1$ and when all model parameters are well constrained with the data. The dashed line is the equality line. Eight dots in the lines indicate the adopted model for the eight GRBs is the single power-law model.}
\label{reduced_chi}
\end{figure}

We finally obtain the following results about the joint spectra of the X-ray flares: Eight flares are adequate to be fitted with the absorbed single power-law model; eleven are well fitted with the absorbed Band function, whereas thirteen are better fitted with the absorbed PL+BB model. To access the adopted model fits, we compare the $\chi^2_r$ of the adopted model fits to that of the single power-law fits in Figure \ref{reduced_chi}. One can find that the absorbed PL+BB model and the absorbed Band function model significantly improve the fits over that with the absorbed PL model for the 2/3 of the spectra in our sample. Note that the $\chi^2_{r}$ of the finally adopted single PL fits of two GRBs (06060607A and 130606A) are larger than 1.1, since the fits with the PL+BB models and the Band function cannot reduce the $\chi^2$. The observed spectra with our fitting curves are shown in Figures \ref{PL}-\ref{PL+BB} and the results are reported in Tables \ref{Table_8PL}-\ref{Table_13BB}. The quoted errors of spectral fitting parameters are in confidence level of 90\%.

\begin{figure}
\centering
\includegraphics[angle=0,scale=0.3]{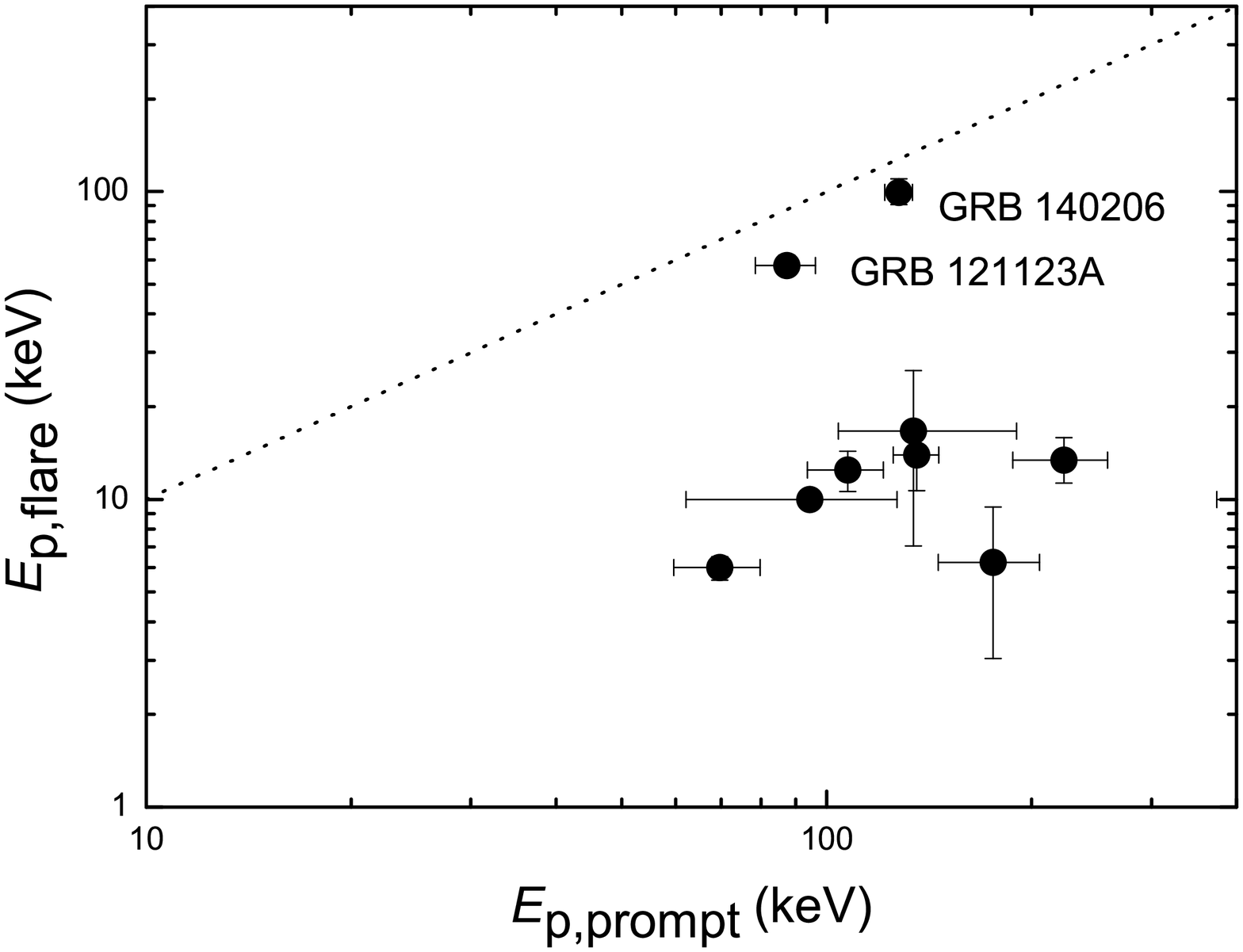}%
\includegraphics[angle=0,scale=0.3]{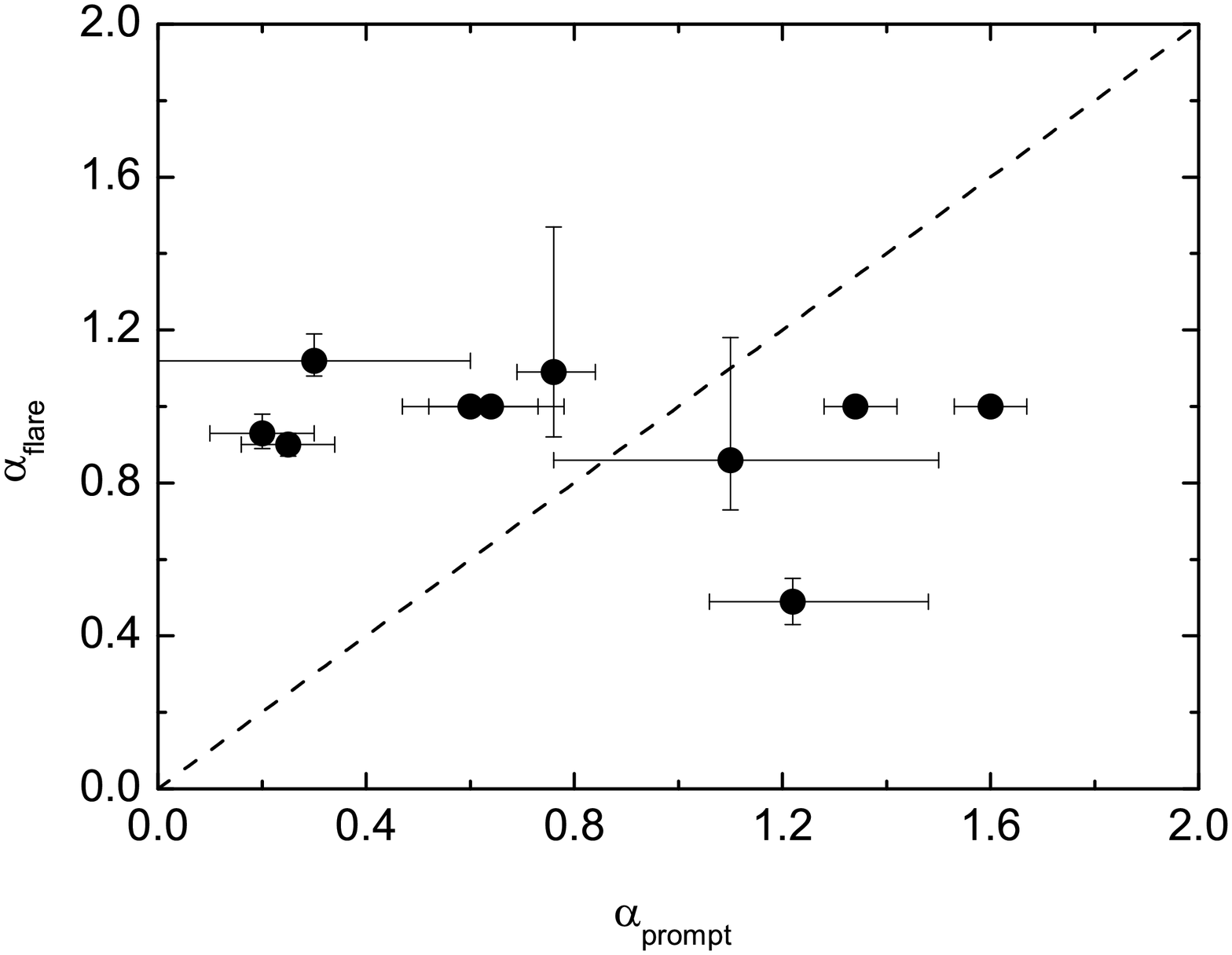}%
\caption{Comparisons of the spectral parameters ($E_p$ and $\alpha$) between the prompt gamma-rays and the flares for 10 GRBs in our sample. The spectra of these flares are fitted with the Band function. These GRBs were also observed with the Fermi/GBM or Konus-Wind. The spectra parameters of the prompt gamma-rays are taken from the Fermi/GBM Catalog or from GCN Circulars (Golenetskii et al. 2009). The dashed lines are the equality line.}
\label{Comp_Parameters}
\end{figure}

The selected X-ray flares in our sample are also detected by BAT, so they also belong the prompt emission phase, even though most of them were detected at the late episode of the prompt emission phase. We compare the spectral parameters between the X-ray flares and the earlier prompt gamma-ray emission before X-ray flares are detected. As is well known, the dominant component in the prompt gamma-ray emission spectra is the Band function component (Band et al. 1993; Zhang et al. 2011). Owing to the narrow energy coverage of BAT, the prompt gamma-ray spectra observed with BAT are usually adequately fitted with a single power-law model. The spectra of the selected flares in 11 GRBs in our sample can be fitted with the Band function. Ten out of these 11 GRBs were also observed with the Fermi gamma-ray burst monitor (GBM) or Konus-Wind during the earlier prompt emission phase. The spectra of these GRBs are well fitted with the Band function or a cutoff power-law models. We collect the spectral parameters of these GRBs from the Fermi/GBM Catalog  (Goldstein et al. 2012;  Paciesas et al. 2012.)\footnote{http://heasarc.gsfc.nasa.gov/W3Browse/fermi/fermigbrst.html.} or from GCN Circulars (Golenetskii et al. 2009). Figure \ref{Comp_Parameters} shows the comparisons of $E_{\rm p}$ and $\alpha$ between the prompt gamma-ray emission (time-integrated spectra including both earlier $\gamma$-rays and later $\gamma$-rays overlapping with X-ray flares) and the X-ray flares\footnote{We do not make a comparison of the high energy spectral index $\beta$, since the spectra of some GRBs are fitted with the cutoff power-law model.}. One can observe that the $E_p$'s of the flares are usually much lower than those of prompt gamma-rays, except for GRBs 121123A and 140206 since the flares in these two GRBs were detected during the peak time of prompt gamma-ray emission. The $\alpha$ values of the flares are typically in the range of 0.8-1.2, whereas it scatters in the range of 0-1.6 for the prompt gamma-rays. Notice that GRB spectra usually show significant evolution with time and the evolution feature among GRBs is diverse (e.g., Liang et al. 1996, Lu et al. 2010, 2012). Thus revealing time-resolved spectral with differe t $E_p$ using overlapping data between BAT and XRT is essential to unveil radiation physics (e.g. Lyu et al. 2014).

\section{Constraints on the physical properties of X-ray flares
with the observed BB component}

 \subsection{Baryonic photosphere model}
As shown in (M\'esz\'aros \& Rees 2000), a thermal component should be embedded in the observed GRB spectrum, if the GRB outflow is matter dominated. For a continuous outflow, the photosphere emission should be also continuous as thermal photons entrained in the outflow progressively become transparent. As a result, early prompt gamma-ray emission and later X-ray flares should have their own photosphere emission components. The thermal component in the BAT-XRT joint spectra would reveal the photosphere properties of the X-ray flares. The same theoretical framework can be applied to study the properties of the X-ray flare photosphere, if the outflow is matter dominated.

In the following, we apply the standard baryonic photosphere model (M\'esz\'aros \& Rees 2000) to constrain the physical properties of the X-ray flare outflows with the detections of a BB component.

Let us assume that a total
luminosity of $L_0$ is released at an initial radius $R_0$, the initial dimensionless entropy is defined as $\eta=L_0/\dot{M}c^2$,
where $\dot{M}$ is the mass rate of the outflow and $c$ is the
speed of light. The initial temperature of the fireball is
\begin{equation}
T_0=(L_0/4\pi R_0^2 a
c\Gamma_i^2)^{1/4}/(1+z)=1.1 {\rm MeV}\ L_{0,52}^{1/4}R_{0,7}^{-1/2}\Gamma_i^{-1/2}/(1+z),
\end{equation}
where $\Gamma_i$ is the initial Lorentz factor at $R_0$, $a$ is
the radiation constant $a=7.57\times 10^{-15}{\rm erg\ cm^{-3}}
K^{-4}$, and the notation $Q_{\rm n}$ is defined as $Q/10^{n}$ in
the cgs units. The scattering optical depth for the fireball with
radius $R$ and Lorentz factor $\Gamma$ is given by
\begin{equation}
\tau=\sigma_T n' \delta R=\sigma_{\rm T} L_0/(4\pi R^2 m_p c^3 \eta
\Gamma)(R/2\Gamma),
\end{equation}
where $\sigma_{\rm T}$ is the Thompson scattering cross section. Taking $\tau=1$, one obtains the photosphere radius as
\begin{equation}
R_{\rm ph}=\frac{\sigma_T L_0}{8\pi m_p c^3 \eta \Gamma_{\rm ph}^2},
\end{equation}
where $\Gamma_{\rm ph}$ is the Lorentz factor of the photosphere,
$m_p$ is the proton mass. Since the photosphere of a baryonic
fireball is usually above the acceleration saturation radius, we
have $\Gamma_{\rm ph}\simeq\eta$, then
\begin{equation}
R_{\rm ph}=\frac{\sigma_{\rm T} L_0}{8\pi m_p c^3 \Gamma_{\rm ph}^3}.
\end{equation}
The luminosity of the BB component is given by
\begin{equation}
L_{\rm BB}=4\pi R^2 \Gamma_{\rm ph}^2 a T'^4 c=4\pi
R^2\frac{(1+z)^4}{\Gamma_{\rm ph}^2}\sigma T^4,
\end{equation}
where  $T'$ and $T=\Gamma_{\rm ph}T'/(1+z)$ are the photosphere
temperature in the comoving frame and the observer frame, respectively.
The blackbody luminosity $L_{\rm BB}$ can be calculated with
the observed flux, i.e.,
$L_{BB}=4\pi D_L^2 F_{\rm BB}$. Therefore, we obtain
\begin{equation}\label{Gamma_ph}
 { \Gamma_{\rm ph}=[\frac{(1+z)^2}{D_{\rm L}}\frac{\sigma_T L_{0}}{8\pi m_p c^3}(\frac{\sigma T^4}{F_{BB}})^{1/2}]^{1/4}}
\end{equation}
and
\begin{equation}\label{R_ph}
 { R_{\rm ph}=[\frac{D_L^3}{(1+z)^6}\frac{\sigma_T  L_{0}}{8\pi m_p c^3}(\frac{F_{BB}}{\sigma T^4})^{3/2}]^{1/4}}.
\end{equation}
Assuming a radiation efficiency of 0.2 and the $k$-correction of a
factor 2 for the non-thermal component with Band function
parameters ($\alpha=-1$, $\beta=-2.3$, and $E_{\rm p}=300$ keV),
we may estimate $L_0\sim 10 L_r$, with $L_r$ dominated by the
non-thermal component. Eight out of the 13 GRBs with the BB component detection have
redshift information. Using the observed flux and
temperature of the BB component, one can obtain $\Gamma_{\rm ph}$
and $R_{\rm ph}$ for the 8 GRBs, which are presented in Table \ref{Table_Fireball}. The derived
$\Gamma_{\rm ph}$ is in the range of $50\sim 150$ and the
photosphere radius is about $R_{\rm ph}\sim 10^{13}{\rm cm}$.
Since the photosphere radius is above the acceleration saturation
radius,  the Lorentz factor of emission region producing X-ray
flares should be nearly equal to $\Gamma_{\rm ph}$. This provides
a unique approach to estimate the Lorentz factor of the X-ray
flares, which is poorly known so far. Since these are bright X-ray
flares, which occur at the late episode of the prompt emission,
it is not unreasonable to have the inferred Lorentz factor $\ga 100$.

In the baryonic jet scenario, the Lorentz factor of the outflow
evolves as
\begin{equation}
\Gamma(R)=\left \{
\begin{array}{ll}
R/R_0,\,\,\, {\rm if}\,\, R<R_{\rm sat}
\\{\eta}, \,\,\,\ \ \ \ \ \  {\rm if}\,\, R>R_{\rm sat}
\end{array}
\right.
\end{equation}
where $R_{\rm sat}=\eta R_0$ is the saturation radius for
acceleration. The observer-frame temperature is given by
\begin{equation}
T=\left \{
\begin{array}{ll}
T_0(R_{\rm ph}/R_{\rm sat})^{-2/3},\,\, {\rm if}\,\, R_{\rm ph}>R_{\rm sat} \\
T_0, \,\,\ \ \ \ \ \ \ \ \ \ \ \ \ \ \ \ \ \ \  {\rm if}\,\, R_{\rm ph}<R_{\rm sat}
\end{array}
\right.
\end{equation}
where $T_{0}$ is the initial temperature of the fireball at $R_{0}$. The photosphere luminosity is expected to be
\begin{equation}
L_{BB}=\left \{
\begin{array}{ll}
L_0(R_{\rm ph}/R_{\rm sat})^{-2/3},\,\, {\rm if}\,\, R_{\rm ph}>R_{\rm sat} \\
L_0, \,\,\ \ \ \ \ \ \ \ \ \  {\rm if}\,\, R_{\rm ph}<R_{\rm sat}
\end{array}
\right.
\end{equation}
Thus, $R_{\rm sat}$, $R_0$, and $T_{0}$ can be estimated with
\begin{equation}
\left \{\begin{array}{ll}
R_{\rm sat}=R_{ph}(L_{\rm BB}/L_0)^{3/2},\\
R_0=R_{\rm sat}/\Gamma_{\rm ph},\\
T_0=T(R_{\rm ph}/R_{\rm sat})^{2/3}=T_{\rm ph}(L_0/L_{\rm BB}).
\end{array}
\right.
\end{equation}
The results are reported in Table \ref{Table_Fireball}. The typical
values of $R_0$ and $R_s$ are $\sim 10^{8}-10^9 {\rm cm} $ and a
few $\sim 10^{10}{\rm cm}$, respectively. Interestingly, the value
of the initial fireball radius $R_0$ is comparable to that derived
from some GRBs using the prompt emission thermal emission data
(e.g. in GRB090902B, Ryde et al. 2010).
These results suggest that at least for some X-ray flares,
emission may be from a baryon-dominated fireball.

\subsection{$\Gamma_{\rm ph}-L_{\rm BB}$ and $E_{\rm p}-L_{\rm BB}$ correlations for the photosphere radiation}
\begin{figure}
\centering
\includegraphics[angle=0,scale=0.5]{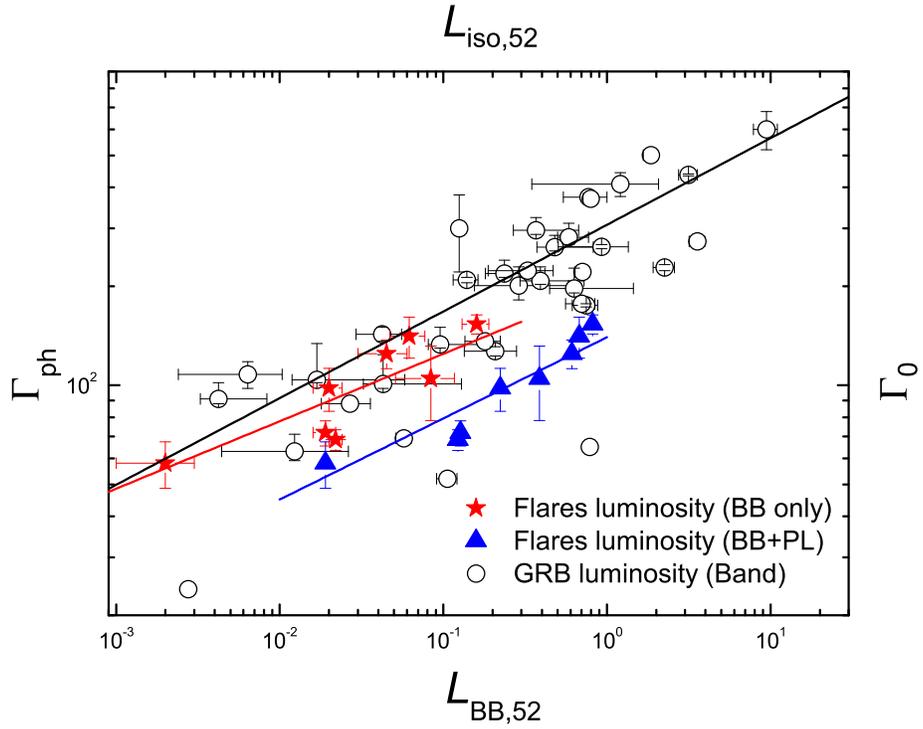}%
\hfill
\caption{$\Gamma_{\rm ph}$ as a function of luminosity for the 8 flares with detection of the BB component and redshift measurement. The red stars denote $L_{\rm BB}$ of the flares, while the blue up-triangles denote the total (thermal plus non-thermal) luminosity in the XRT+BAT band. The $\Gamma_0-L_{\gamma, \rm iso}$ relation for GRBs in general (L\"{u} et al. 2012) are over-plotted (open circle) for comparison. The solid lines are the best fit to the data.}
\label{Gamma_LBB}
\end{figure}

Liang et al. (2010) and L\"{u} et al. (2012) found tight correlations
between the initial  Lorentz factor and the isotropic gamma-ray
energy/luminosity. With the $\Gamma_{\rm ph}$ values derived above
for the 8 flares, we show $\Gamma_{\rm ph}$ as a function of
$L_{\rm BB}$ in comparison with the $\Gamma_0-L_{\rm \gamma, iso}$
in Figure \ref{Gamma_LBB}. Interestingly, a tight $\Gamma_{\rm ph}-L_{\rm BB}$ relation is found, i.e., $\log \Gamma_{\rm ph}=(2.30\pm 0.08)+(0.20\pm 0.05)\log L_{\rm BB, 52}$, with a
Spearman correlation coefficient of 0.69 and a chance probability 0.006 ($N=8$). Adding the non-thermal component to the luminosity, the relation becomes $\log \Gamma_{\rm ph}=(2.15\pm 0.03)+(0.25\pm 0.03)\log L_{\rm BB+PL, 52}$, with a
Spearman correlation coefficient of 0.88 and a chance probability
$p\sim 10^{-4}$ ($N=8$). We also over-plot the $\Gamma_0-L_{\rm \gamma,iso}$ relation (L\"{u} et al. 2012) for comparison. Note that the $\Gamma_0$ reported in L\"{u} et al. (2012) were estimated using other methods. In particular, for most GRBs, $\Gamma_0$ is estimated using the fireball deceleration timescales measured in the early afterglow lightcurves, and hence, stands for the Lorentz factor before the deceleration time. On the other hand, as mentioned above, the baryonic photosphere radius is larger than the saturation radius $R_{\rm sat}$. Therefore, the $\Gamma_{\rm ph}$ is essentially comparable to $\Gamma_0$. It is found that the  $\Gamma_{\rm ph}-L_{\rm BB}$ relation is well consistent with $\Gamma_0-L_{\rm \gamma,iso}$ relation.

The physical origin of $\Gamma_0-L_{\rm \gamma,iso}$ is
unknown. The observed $\Gamma_{\rm ph}-L_{\rm BB}$
relation is a natural consequence of the baryonic photosphere model. Within this model, one expects $\Gamma_{\rm ph} \propto L_{\rm BB}^{0.27} R_0^{-0.18}$. If $R_0$ does not vary significantly within a burst, the correlation
between $\Gamma_{\rm ph}$ and $L_{\rm BB}$ is naturally expected. Fan et al. (2012) started from
radiation physics and suggested that emission from a baryonic photosphere
can naturally produce a $\Gamma_{\rm ph} - L_{\rm BB}$ relation and suggested that the observed $\Gamma_0-L_{\rm \gamma,iso}$ can be attributed
to this, suggesting that the bulk of GRB emission comes from the baryonic
photosphere (see also Lazzati et al. 2013).

The consistency between the $\Gamma_{\rm ph}-L_{\rm BB}$
relation and the $\Gamma_0-L_{\rm \gamma,iso}$
relation (L\"u et al. 2012) is intriguing. A direct implication would
be that the prompt emission of GRBs is photosphere-dominated, as
suggested by Fan et al. (2012). Indeed some GRBs (e.g. GRB 090902B
and GRB 081221) have been shown to be thermal-dominated (Ryde et al.
2010; Zhang et al. 2011; Hou et al. 2014). These GRBs indeed follow
the $\Gamma_0-L_{\rm \gamma,iso}$ relations. On the other hand,
the dominant emission component of most GRBs, namely, the traditional
``Band'' component, is likely not of a thermal origin. The arguments
against the thermal origin of the Band-function peaks include
the following: 1. The observed spectral index below $E_{\rm p}$ is too
soft to be interpreted by the photosphere model (Deng \& Zhang 2014);
2. The hard-to-soft evolution across broad pulses as observed in
many GRBs (Lu et al. 2010, 2012) is difficult to interpret within
the photosphere model (Deng \& Zhang 2014) but is straightforward
to interpret within the synchrotron emission model (Uhm \& Zhang
2014; Bosnjak \& Daigne 2014); 3. The $E_{\rm p}$ of the Band component
sometimes is above the ``death line'' defined by the photosphere
model (Zhang et al. 2012); 4. Most importantly, several GRBs
(100724B, 110721A, 120323A, Guiriec et al. 2011; Axelsson et al.
2012; Guiriec et al. 2013) have both the photosphere component
and the Band component identified, with the photosphere component
being the sub-dominant component. The X-ray flares in our sample, even if with a thermal
component in the spectrum, are also non-thermal dominated.
Adding the non-thermal component to the luminosity, these
flares also roughly align with the $\Gamma_0-L_{\rm \gamma,iso}$
relation, even though introducing some scatter to the
correlation. It was also proposed that the observed
$\Gamma_0-L_{\rm \gamma,iso}$ correlation in GRBs may stem from an intrinsic physical reason related to central engine (Lei et al. 2013), who showed that for a hyper-accreting black hole central engine both a neutrino-anti-neutrino annihilation model (which produces
a fireball) and a Blandford-Znajek model (which produces a magnetically
dominated outflow) can give rise to a $\Gamma_0-L_{\rm \gamma,iso}$
correlation similar to what is observed.
We note that the $\Gamma_{\rm ph}-L_{\gamma, \rm iso}$ relation of
X-ray flares (blue triangles) in Fig.\ref{Gamma_LBB} lines below the
main correlation of prompt gamma-rays. This may be a hint that the
outflow is more magnetically dominated. Indeed, a generalized photosphere
model invoking an arbitrary magnetization of the outflow (Gao \& Zhang 2014,
see also Veres \& M\'esz\'aros 2012)
suggests that the photosphere usually occurs during the phase when the
outflow is still being accelerated. Also, the thermal to non-thermal
flux ratio is low due to the early suppresion of photosphere emission
from an magnetized outflow and later efficient release of magnetic energy
due to ICMART processes (Zhang \& Yan 2011).

\begin{figure}
\centering
\includegraphics[angle=0,scale=0.5]{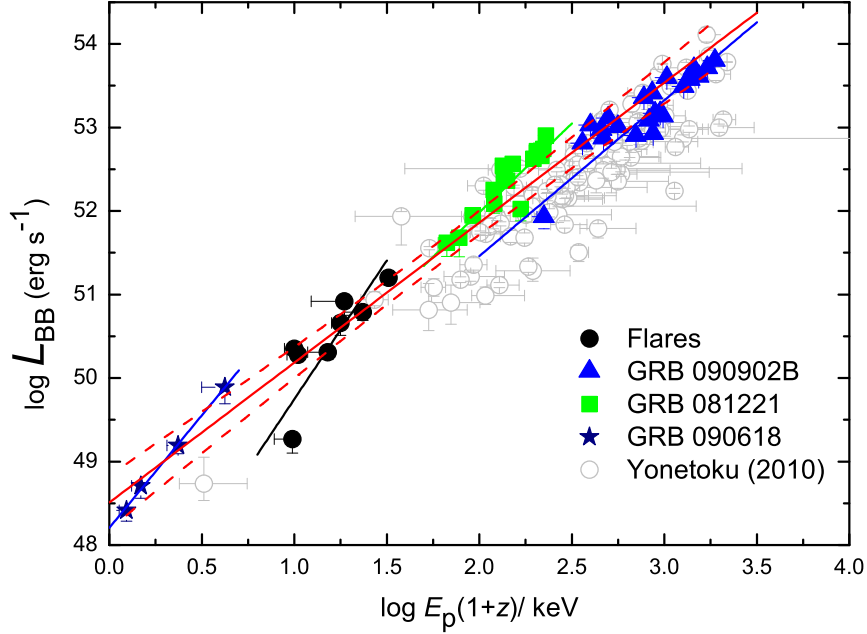}%
\hfill
\caption{Luminosity as a function of the $E_{\rm p}$ of the BB component for 8 GRBs with detection of the BB component and redshift measurement. Time-resolved $L_{\rm BB}-E_{\rm p}$ relation for the BB component observed in the flares of GRBs 081221, 090618, and 090902B are also shown. The Yonetoku relation for the Band function component in typical GRBs is over-plotted for comparison with a sample from Yonetoku et al. (2010). Solid lines are the best fit to the data. The best linear fit and its 3$\sigma$ confidence level for the data points of flares in the 8 GRBs and in GRBs 081221, 090618, and 090902B are also present (red lines). }
\label{Ep_LBB}
\end{figure}

We further explore the $E_{\rm p}-L_{\rm BB}$ relation for  the BB
component. We show $L_{\rm BB}$ as a function of $E_{\rm p}$ for
the flares in comparison with the same correlation
within GRBs 081221 (Hou et al.
2014, in preparation), 090618 (Page et al. 2011)\footnote{The data
of the late time slice 275-2453s of 090618 reported in Page et al.
(2011) is not included since it is not in the prompt phase and the
reduced $\chi^2$ of the fit with a single power-law is 1.08, which
is got enough to fit the data.}, and 090902B (Zhang et al. 2011),
the three GRBs with time resolved thermal emission identified.
Such an $E_{\rm p}-L_{\rm BB}$ correlation is found among flares and
also within individual flares. Generally, the $E_{\rm p}-L_{\rm BB}$
correlation within a GRB is tighter than that among the flares.
Our Spearman correlation analysis yields
$\log L_{\rm
BB}=(46.43\pm 0.95)+(3.32\pm 0.78)\log E^{'}_{\rm p}$ with a
scatter 0.37 for the flares ($r=0.71$, $p=0.005$),
$\log L_{\rm
BB}=(47.72\pm 0.52)+(2.13\pm 0.25)\log E^{'}_{\rm p}$ with a
scatter 0.16 for GRB 081221 ($r=0.81$, $p<10^{-4}$),
$\log L_{\rm
BB}=(47.73\pm 0.45)+(1.87\pm 0.16)\log E^{'}_{\rm p}$ with a
scatter 0.21 for GRB 090902B ($r=0.86$, $p<10^{-4}$),
and $\log L_{\rm
BB}=(48.21\pm 0.04)+(2.70\pm 0.09)\log E^{'}_{\rm p}$ with a
scatter 0.03 for GRB 090618 ($r=0.99$, $p=0.001$). This is
probably because $r_0$ has a small scatter within a same GRB.
The slope ($\rho$) of the $E_{\rm p}-L_{\rm BB}$ correlations (i.e. $L_{\rm
BB}\propto E_{\rm p}^{\rho}$) within individual GRBs varies slightly
among GRBs.
Globally, one can observe an $E_{\rm
p}-L_{\rm BB}$ correlation for all data in Figure \ref{Ep_LBB} in a broad
$E_{\rm p}$ and $L_{\rm BB}$ range. Our best fit gives $\log L_{\rm
BB}=(48.51\pm 0.10)+(1.68\pm 0.05)\log E^{'}_{\rm p}$ ($r=0.95$,
$p<10^{-4}$), with a scatter of 0.28.

When overplotting the $E_{\rm p}-L_{\gamma, \rm iso}$ relation
(Yonetoku et al. 2004; Liang et al. 2004) in Fig.\ref{Ep_LBB},
we find that the $E_{\rm p}-L_{\rm BB}$ relations also align with
the observed $E_{\rm p}-L_{\gamma, \rm iso}$ relation. A straightforward
inference would be that GRB emission is photosphere dominated
(Fan et al. 2012). However, due to the reasons discussed above,
one may not draw this conclusion directly. Indeed, it was found
that there is a correlation between the peak energy of the
thermal and non-thermal components in GRBs (Burgess et al. 2014).
As a result, the non-thermal component would also follow the
Yonetoku relation. Indeed, $E_{\rm p} \propto L^{1/2}$ is naturally
predicted in synchotron radiation models (Zhang \& M\'esz\'aros
2002), especially if the emission radius is not sensitively
dependent on the bulk Lorentz factor (e.g. Zhang \& Yan 2011).
The observed Yonetoku relation may be a consequence of that
both thermal emission and synchrotron non-thermal emission
produce a positive relation between $E_{\rm p}$ and $L$.

\section{Conclusions and Discussion}
We have presented a joint spectral analysis of 32 GRB X-ray
flares that  were simultaneously observed with BAT and XRT. Our
results show a diverse range of flare spectra.
The joint spectra of 8 flares are adequate to
fit with an absorbed single power-law model.The derived
photon indices are $\Gamma_{\rm PL}\ < 2$, which suggests
that their $E_{\rm p}$ of the $\nu F_\nu$ spectra are beyond the XRT+BAT
band. The joint spectra of 11 flares are well fitted with an
absorbed Band function, with an $E_{\rm p}$ in the $3\sim$ 100 keV
range. More interestingly, the joint spectra of 13 X-ray flares are fitted
with the absorbed BB+PL model. The derived temperature is around $1\sim 3$ keV.
Phenomenally, the observed spectra of the 13 GRBs are analogous to
several GRBs whose thermal component was identified in their spectra,
but only with a much lower
temperature. Assuming that the thermal emission is the
photosphere emission of a relativistic fireball, we derive the physical
properties of the 8 flares that have redshift measurements. The
derived $\Gamma_{\rm ph}$ is in the range of $50\sim 150$, and the
typical $R_0$, $R_s$, and $R_{\rm ph}$ are $\sim 10^{8}$, $\sim
10^{10}$, and $\sim 10^{13}$ cm, respectively.
We also find tight correlations for
$\Gamma_{\rm ph}-L_{\rm BB}$ and $E_{\rm p}-L_{\rm BB}$
in the sample, which is straightforwardly within the photosphere
model.

Combining our sample of X-ray flares with GRBs with a detected
thermal component, one may come up with the following overall
picture of GRB thermal emission.
Some GRBs, examplified by GRB 090902B, have a dominant photosphere
emission component with a temperature around 200 keV.
It has an extra non-thermal power-law component in the GBM
and LAT band,
with a peak and turnover energy beyond the LAT band.
The high energy component is likely of an inverse Compton
scattering origin (Pe'er et al. 2012; Beloborodov et al. 2014).
A similar high energy component was also observed in GRB 090926A,
with the high energy peak identified (Ackermann et al. 2011).
Next, in the case of GRB 081221, the time-resolved spectra
show a clear two-component feature, with the lower and higher
peaks having a thermal and non-thermal origin, respectively.
More common cases are the ones having a sub-dominant thermal
emission component, such as
GRBs 090618 (Page et al. 2011), 110721A (Axelsson et al. 2012), 100724B
(Guiriec et al. 2011) and GRB 120323A (Guiriec et al. 2012), in
which the BB component contributes to the total flux with a
fraction of below $10\%$. The BB component in the 13 flares
studied here may contribute to an even lower fraction to the total
flux. The power-law indices of the non-thermal component in these
flares are typically $\Gamma < 2$,  suggesting that the peak of this
component may be beyond the BAT band. All these make a sequence of
thermal-component dominance in the
GRB (or X-ray flare) spectra, which may suggest that the GRB
central engine may have a distribution of the initial magnetization
$\sigma$, which causes the diversity of the observed spectra
(e.g. Zhang 2011). The results presented here also suggest
that the non-detection of thermal component from
GBM data in some GRBs may be due to that the thermal component
is weak and soft which can be detected only in the X-ray band.
For 13 bursts with PL plus BB spectral, BAT observation is dominated by non-thermal emission. Meanwhile the mechanism to produce non-thermal emission and its efficiency are not clear, so our results are reasonable in jets with different initial dimensionless entropy $\eta$ and magnetization factor $\sigma_{0}$ (Zhang 2014). A more general theory of photosphere model invoking arbitrary magnetization (Gao \& Zhang 2014) should be applied to fully diagnose the outflow parameters.

Our analysis suggests that the $\Gamma_{\rm ph}-L_{\rm BB}$ and
$E_{\rm p}-L_{\rm BB}$ relations discovered for the X-ray flare
photosphere component seems to be consistent with the observed
$\Gamma_{0}-L_{\gamma, \rm iso}$ and $E_{\rm p}-L_{\gamma, \rm
iso}$ correlations derived from prompt
gamma-ray emission (L\"u et al. 2012; Yonetoku et al. 2002).
These correlations are expected, since they are inherited from
the baryonic photosphere model itself (Fan et al. 2012). On
the other hand, the consistency with the correlations found
in GRBs in general is intriguing.
A straightforward inference would be that the GRB prompt emission
is thermal dominated, and the Band function is modified thermal
emission from the photosphere (Fan et al. 2012; Lazzati et al.
2013). However, in view of the difficulties of the photosphere
model to interpret the Band function (e.g. Deng \& Zhang 2014;
Zhang et al. 2012) and the fact that a dominant non-thermal
component has been discovered in many GRBs (Guiriec et al.
2011,2013; Axelsson et al. 2012), one has to attribute those
apparent consistencies to more profound physical reasons.
For the $\Gamma_{0}-L_{\gamma, \rm iso}$ relation, it has
been found that a hyper-accreting black hole central engine
can naturally produce a similar correlation for both a
thermal fireball and a magnetically dominated jet (Lei et
al. 2013). For the $E_{\rm p}-L_{\gamma, \rm iso}$ relation,
both photosphere emission (Thompson et al. 2007; Fan et
al. 2012; Lazzati et al. 2013) and synchotron radiation
(Zhang \& M\'esz\'aros 2002; Zhang \& Yan 2011) can
reproduce a similar correlation. It is possible that
different energy dissipation mechanisms and radiation
mechanisms by coincidence can produce similar correlations
as observed. Combined the above results: the temperature and flux of  BB component, the dynamic quantities, the relationship $\Gamma_{\rm ph}-L_{\rm BB}$ and  $E_{\rm p}-L_{\rm BB}$, show that the GRB central engine is a hybrid system with both thermal energy and magnetic energy. Maybe for some X-ray flares, emission is from baryon dominated fireball. More data and more detailed analysis of these correlations by separating the thermal and non-thermal components
are needed to better understand the underlying physics of
GRB prompt emission and X-ray flares.

\section*{Acknowledgments}
We appreciate valuable comments from the referee. We thank Bin-Bin Zhang, Kim Page and He Gao for helpful discussion. This work made use of data supplied by the UK Swift Science Data Center at the University of Leicester. It is supported by the National Basic Research Program (973 Programme) of China (Grant 2014CB845800), the National Natural Science Foundation of China (Grants 11025313, 11163001, 11363002), Guangxi Science Foundation (2013GXNSFFA019001), Key Laboratory for the Structure and Evolution of Celestial Objects of Chinese Academy of Sciences, and the Strategic Priority Research Program "The Emergence of Cosmological Structures" of the Chinese Academy of Sciences, Grant No. XDB09000000.

\clearpage

\begin{figure}
\includegraphics[angle=0,scale=0.3]{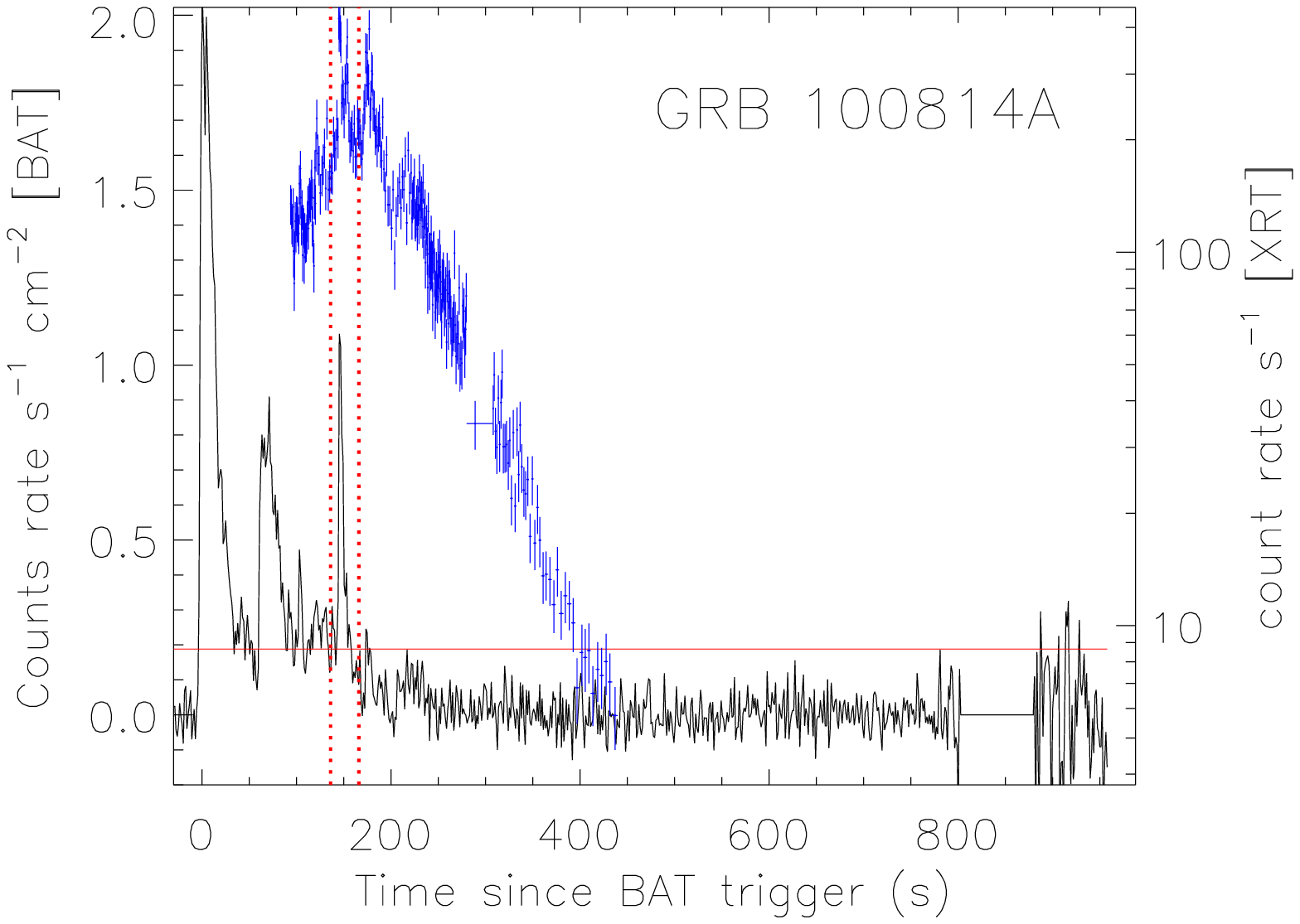}%
\includegraphics[angle=0,scale=0.3]{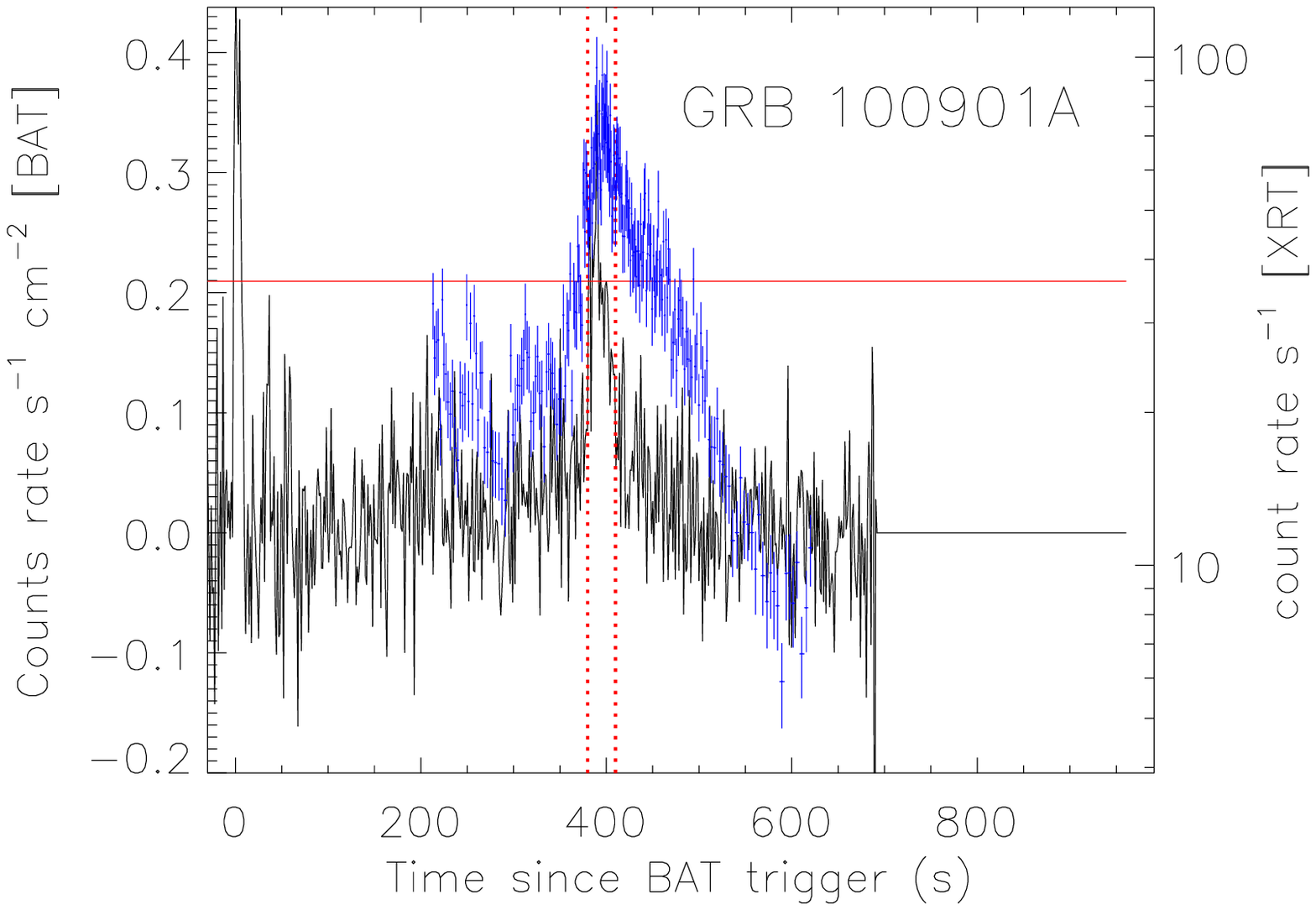}%
\includegraphics[angle=0,scale=0.3]{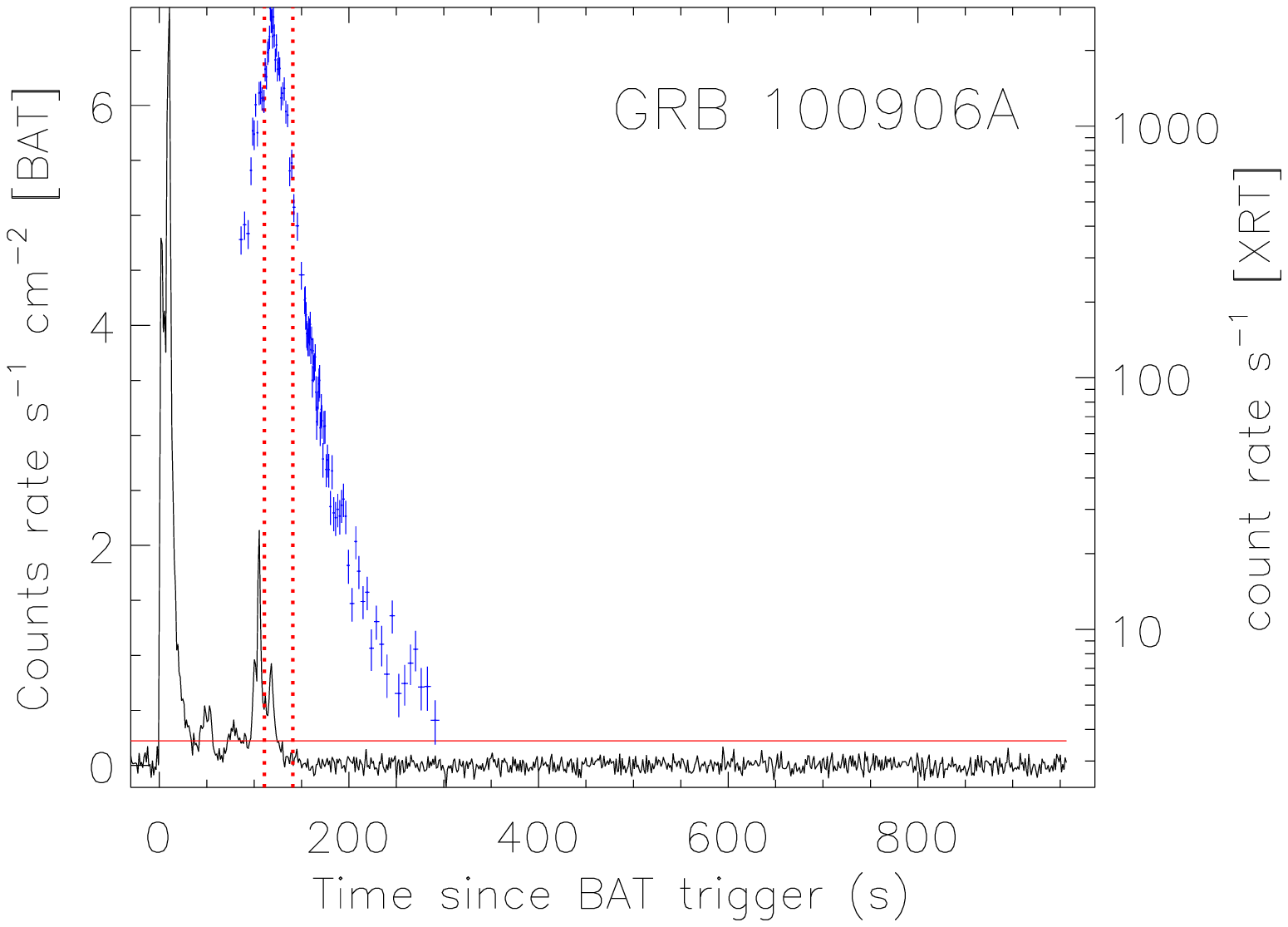}\\
\includegraphics[angle=0,scale=0.3]{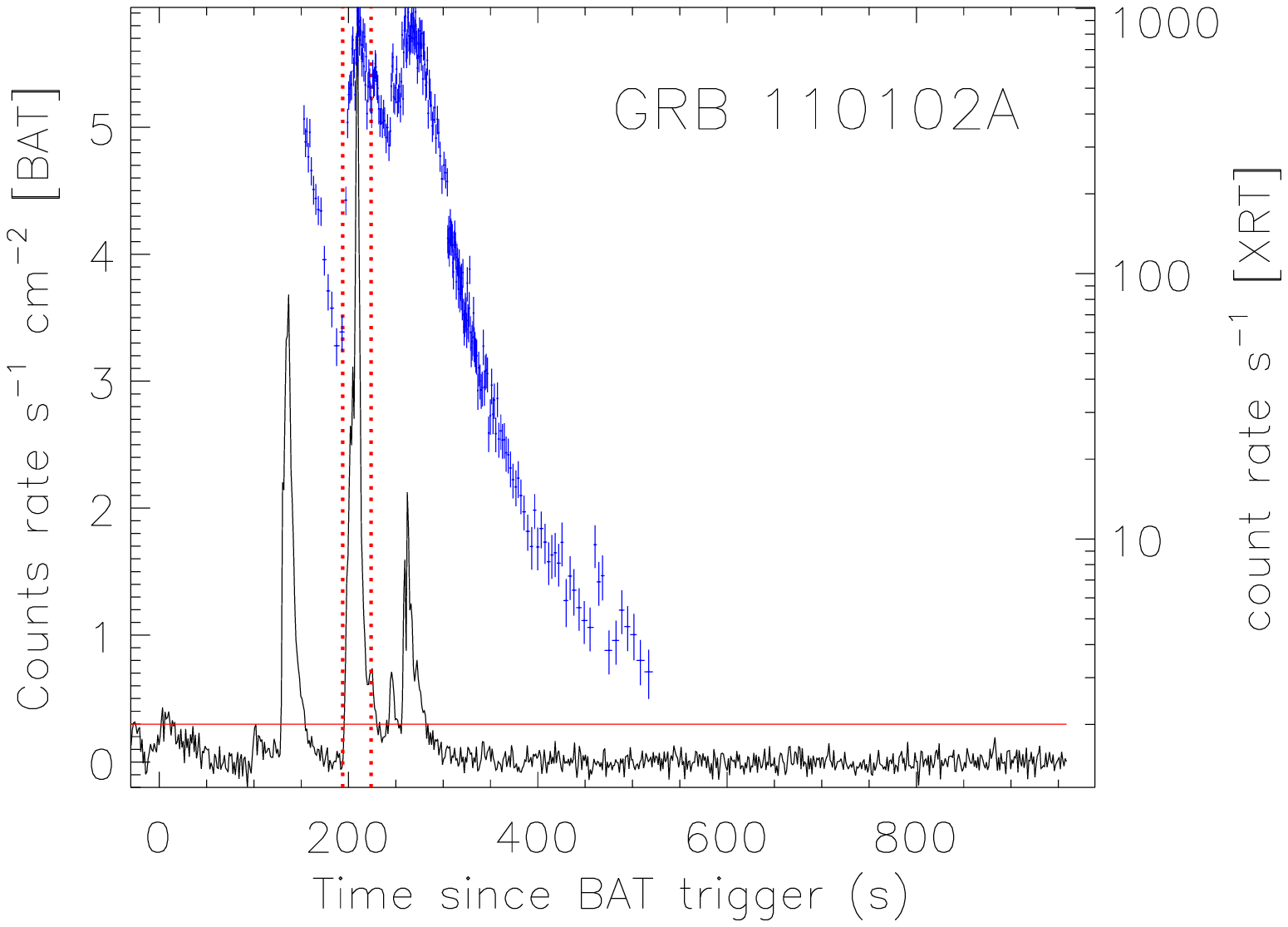}%
\includegraphics[angle=0,scale=0.3]{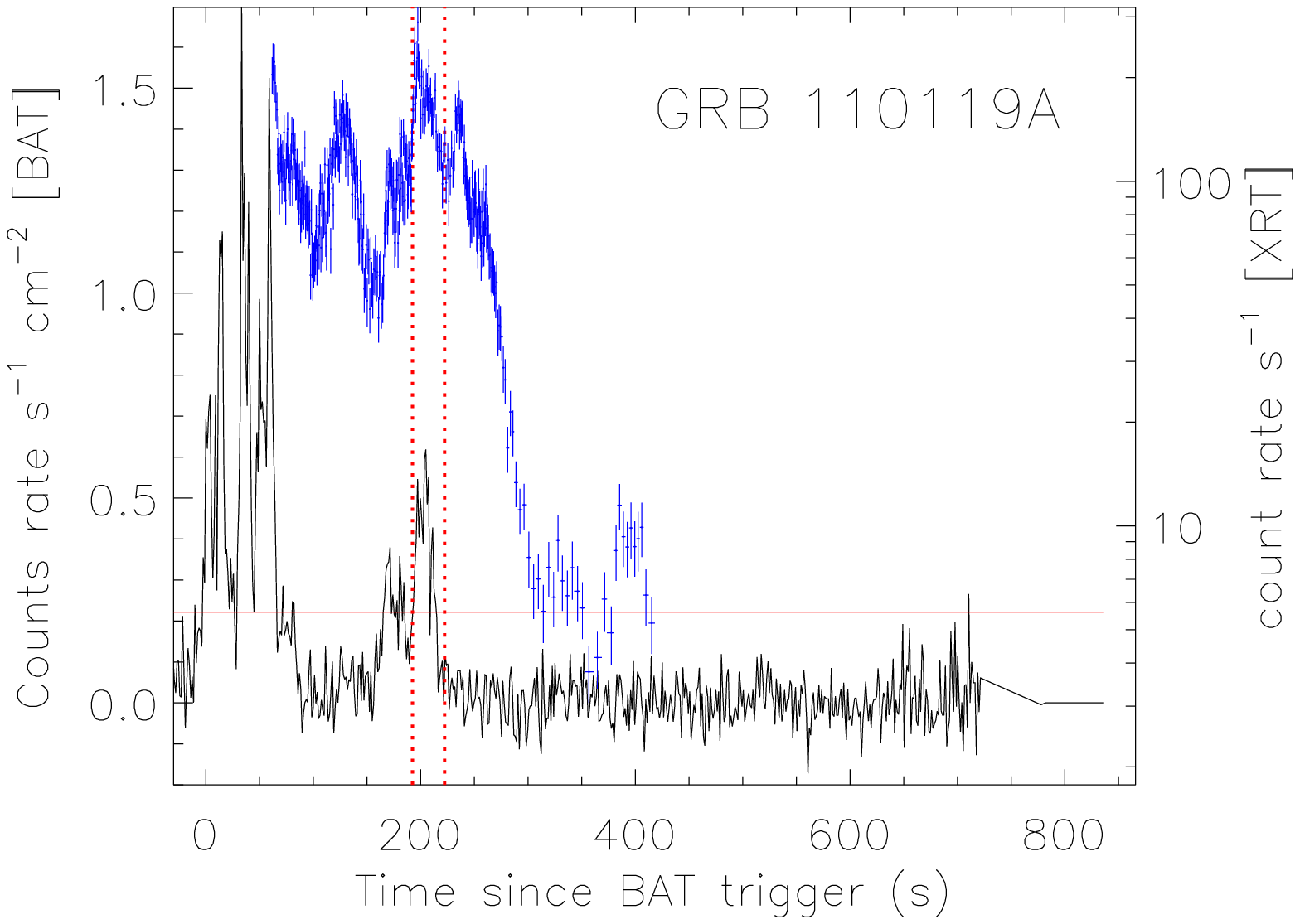}%
\includegraphics[angle=0,scale=0.3]{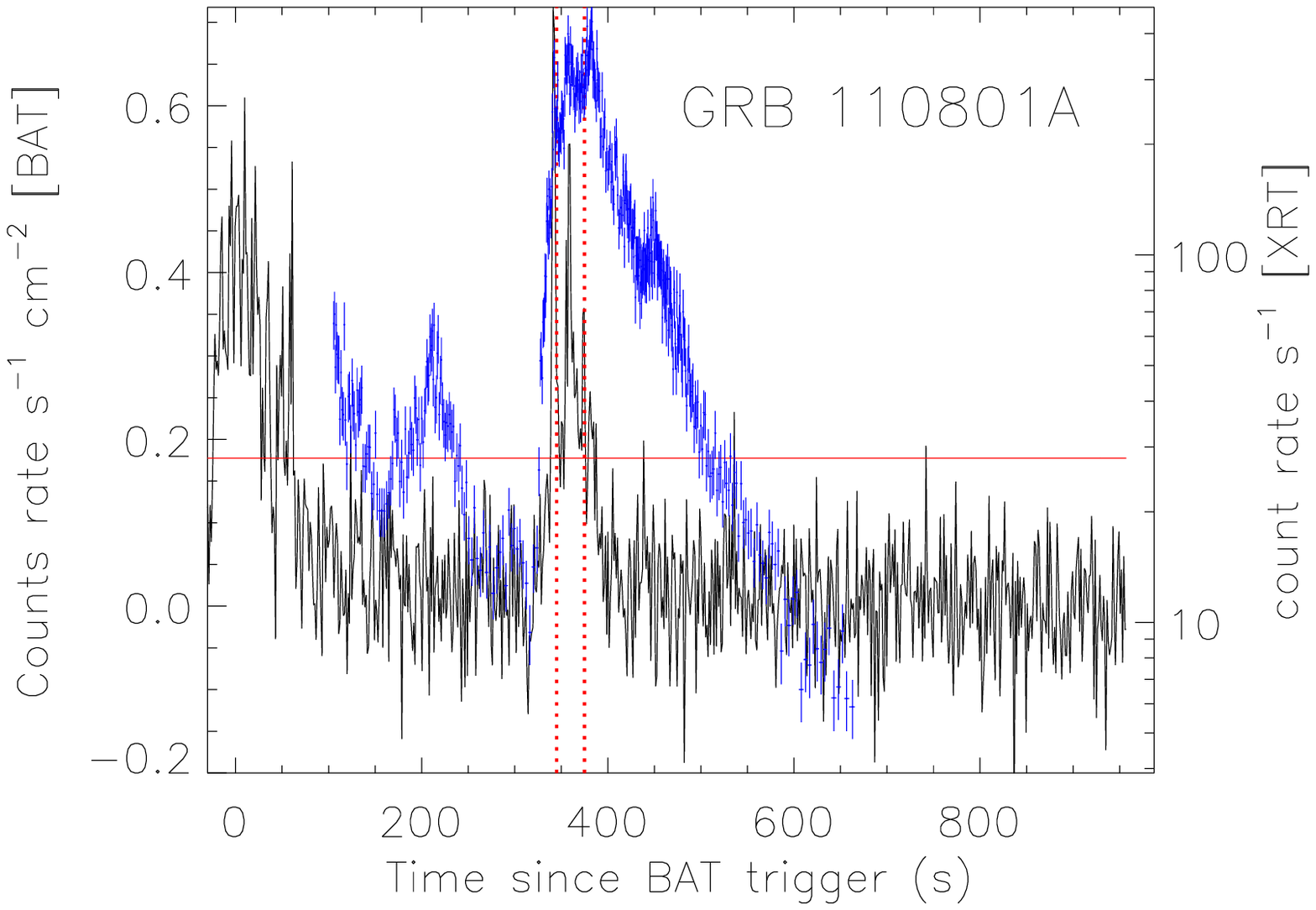}\\
\includegraphics[angle=0,scale=0.3]{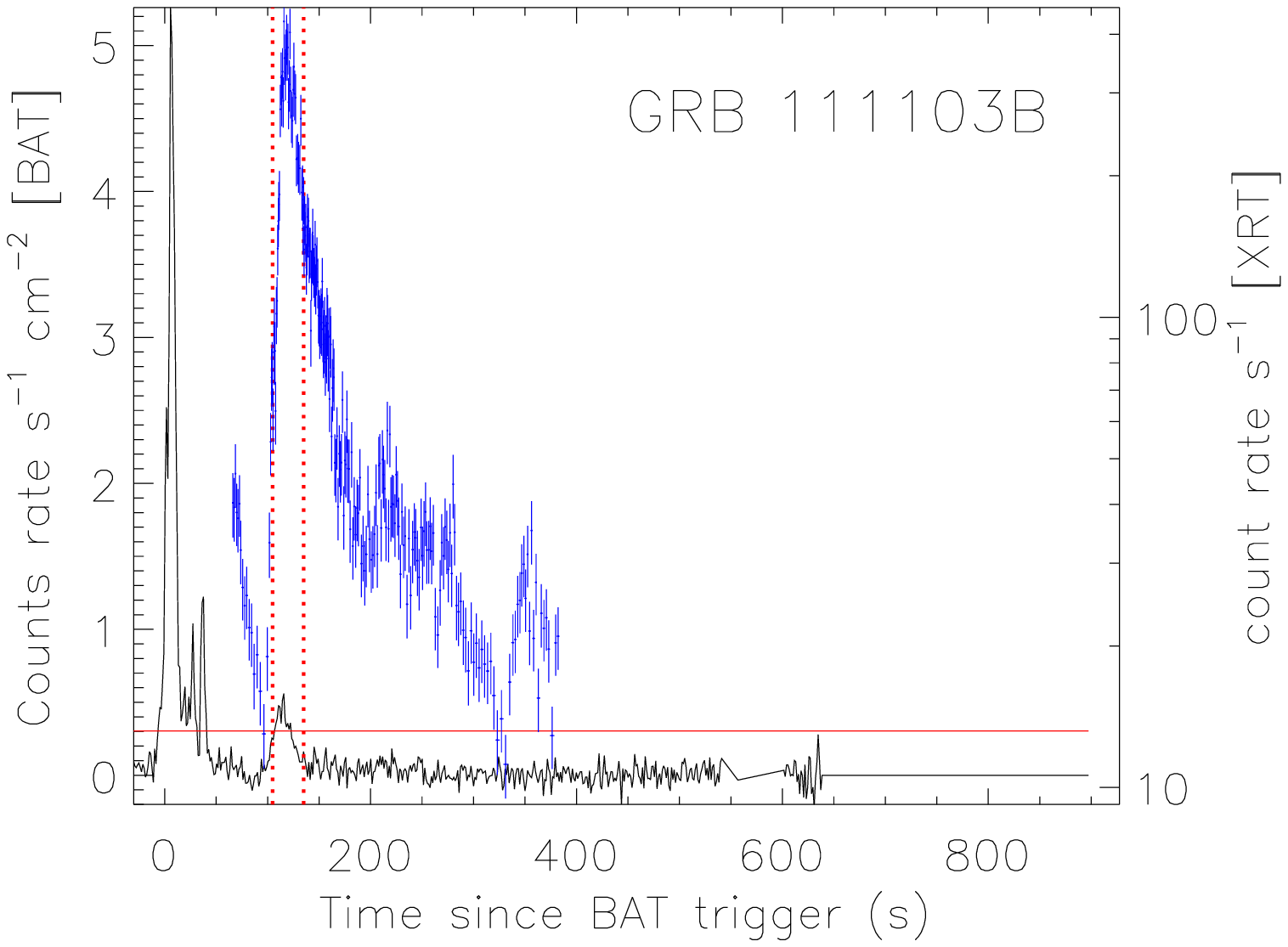}%
\includegraphics[angle=0,scale=0.3]{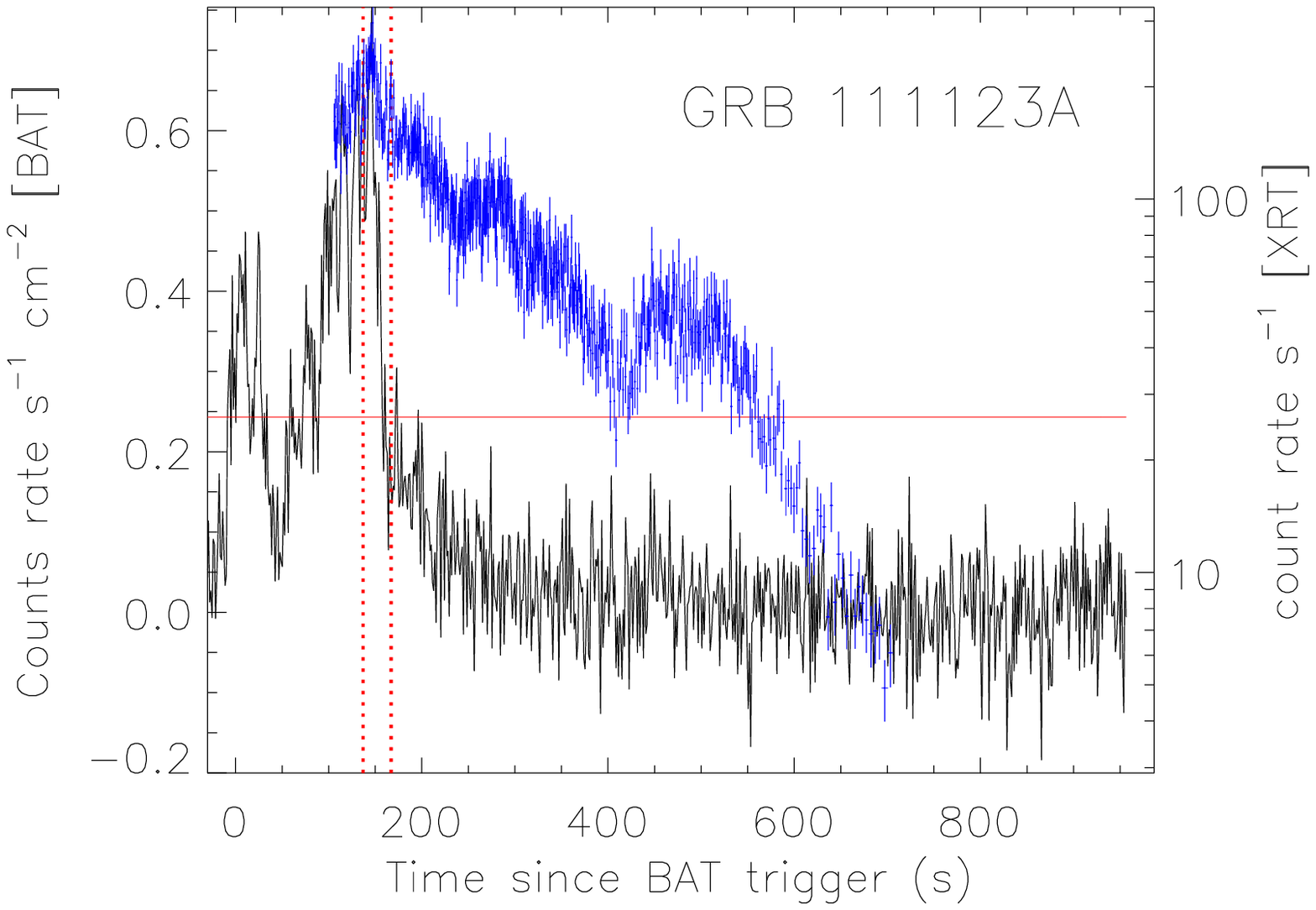}%
\includegraphics[angle=0,scale=0.3]{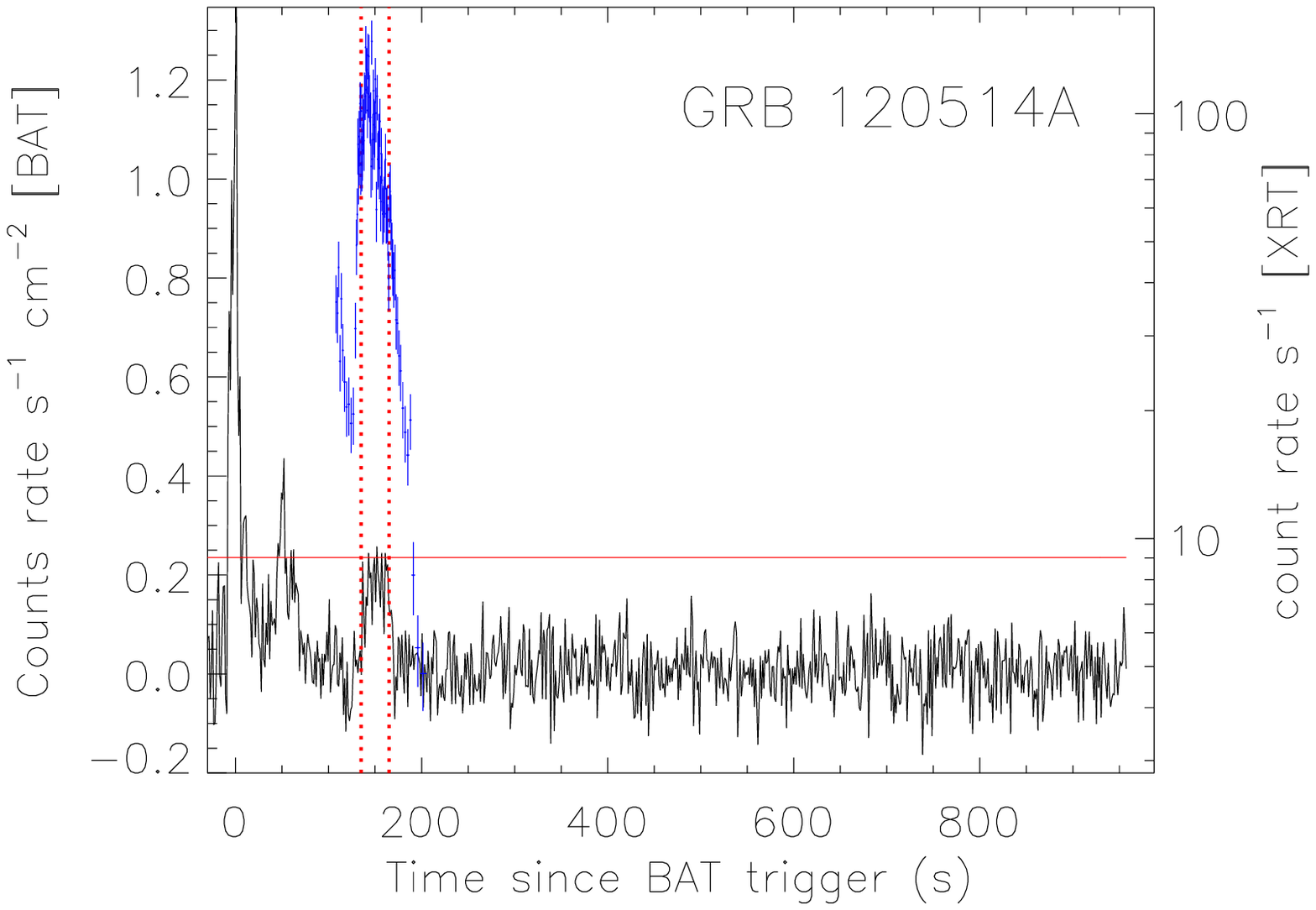}\\
\includegraphics[angle=0,scale=0.3]{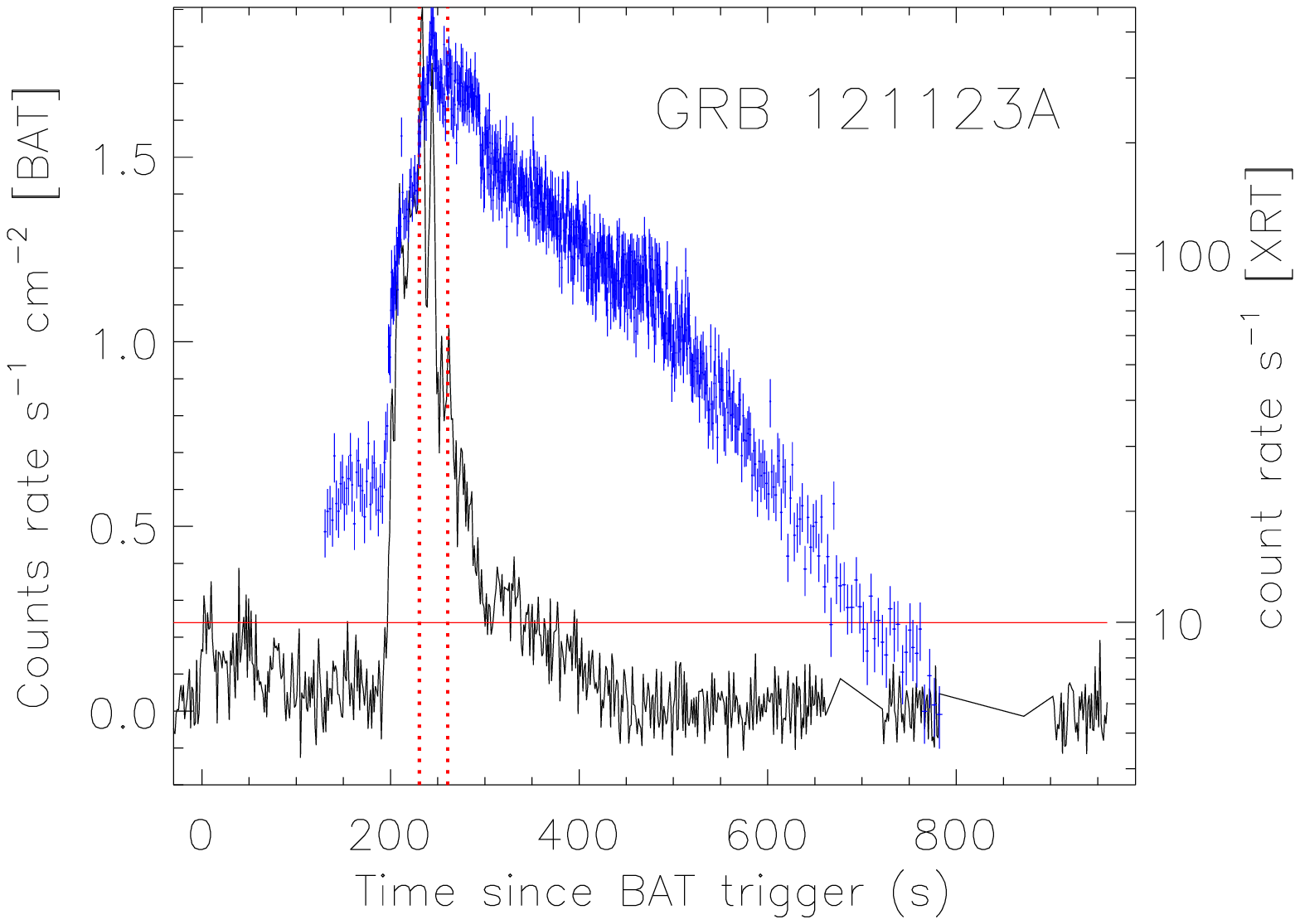}%
\includegraphics[angle=0,scale=0.3]{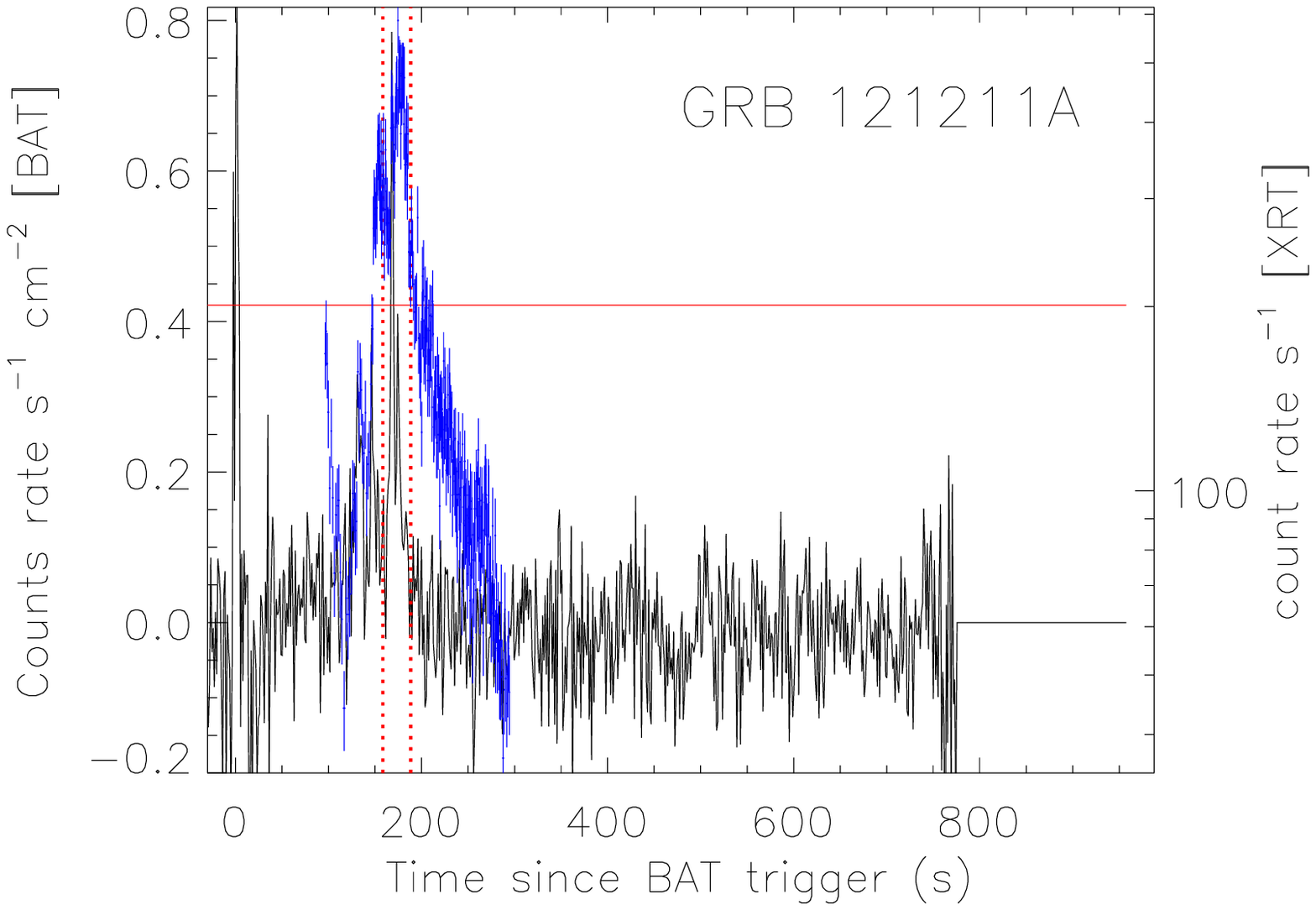}%
\includegraphics[angle=0,scale=0.3]{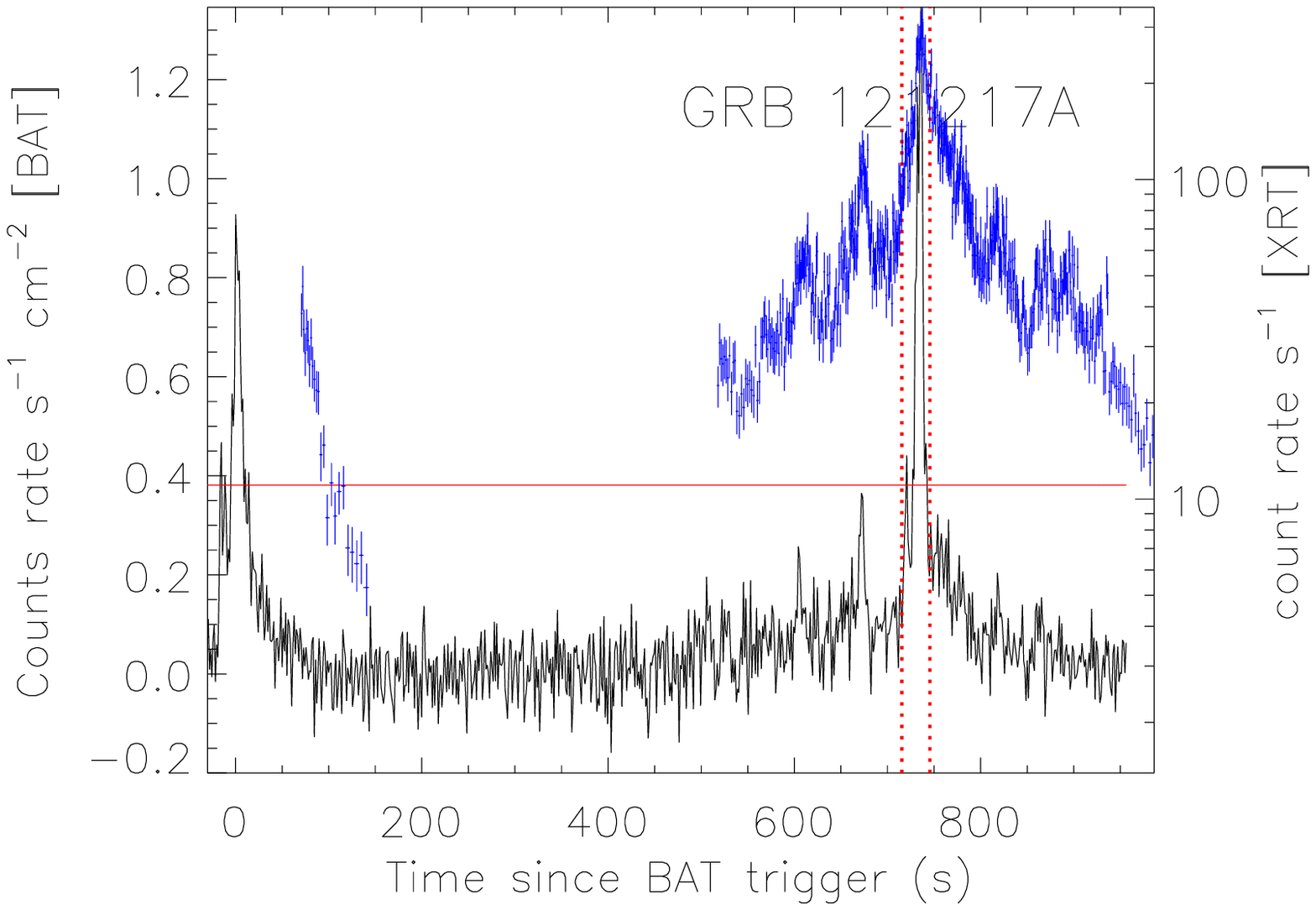}\\
\includegraphics[angle=0,scale=0.3]{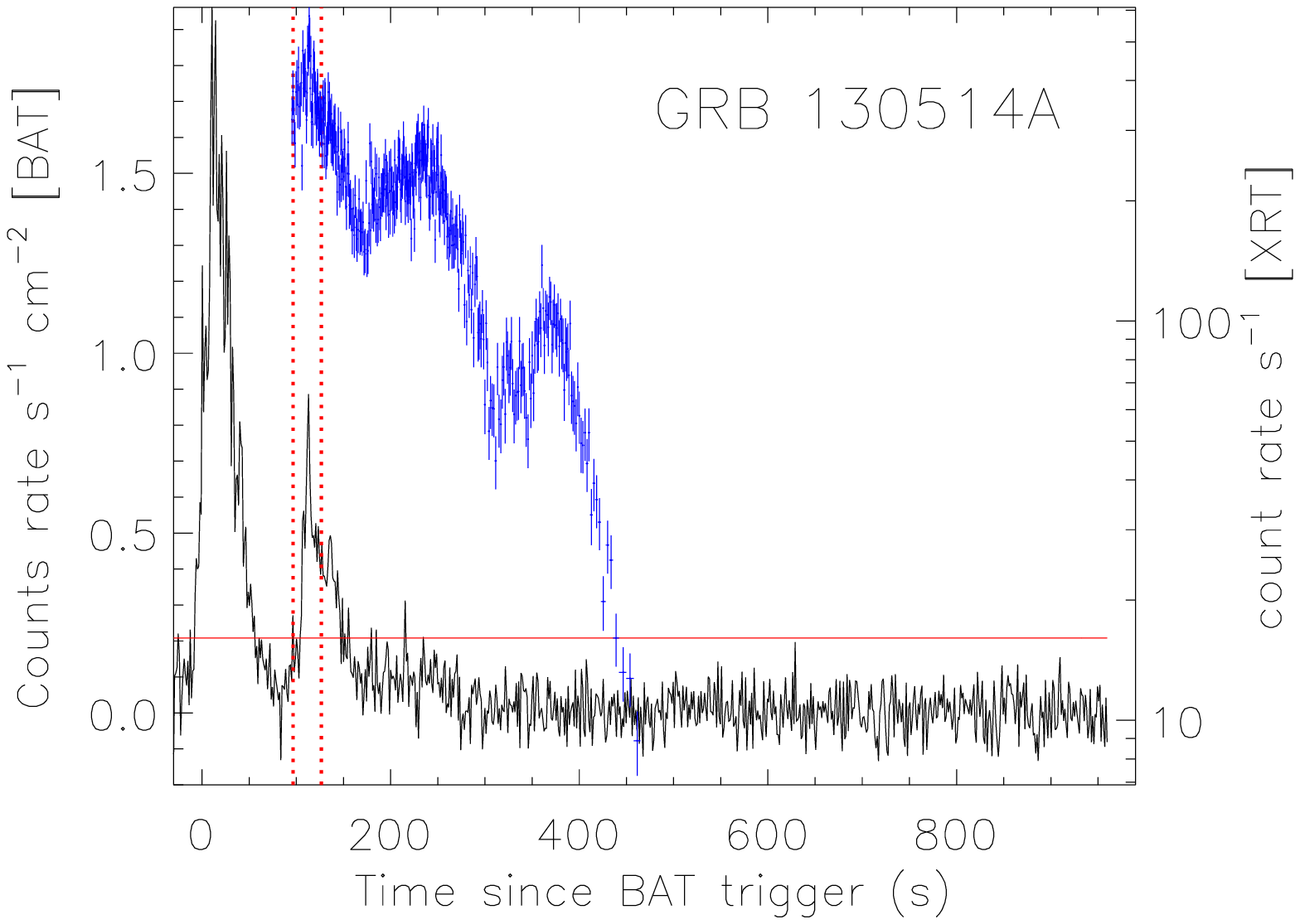}%
\includegraphics[angle=0,scale=0.3]{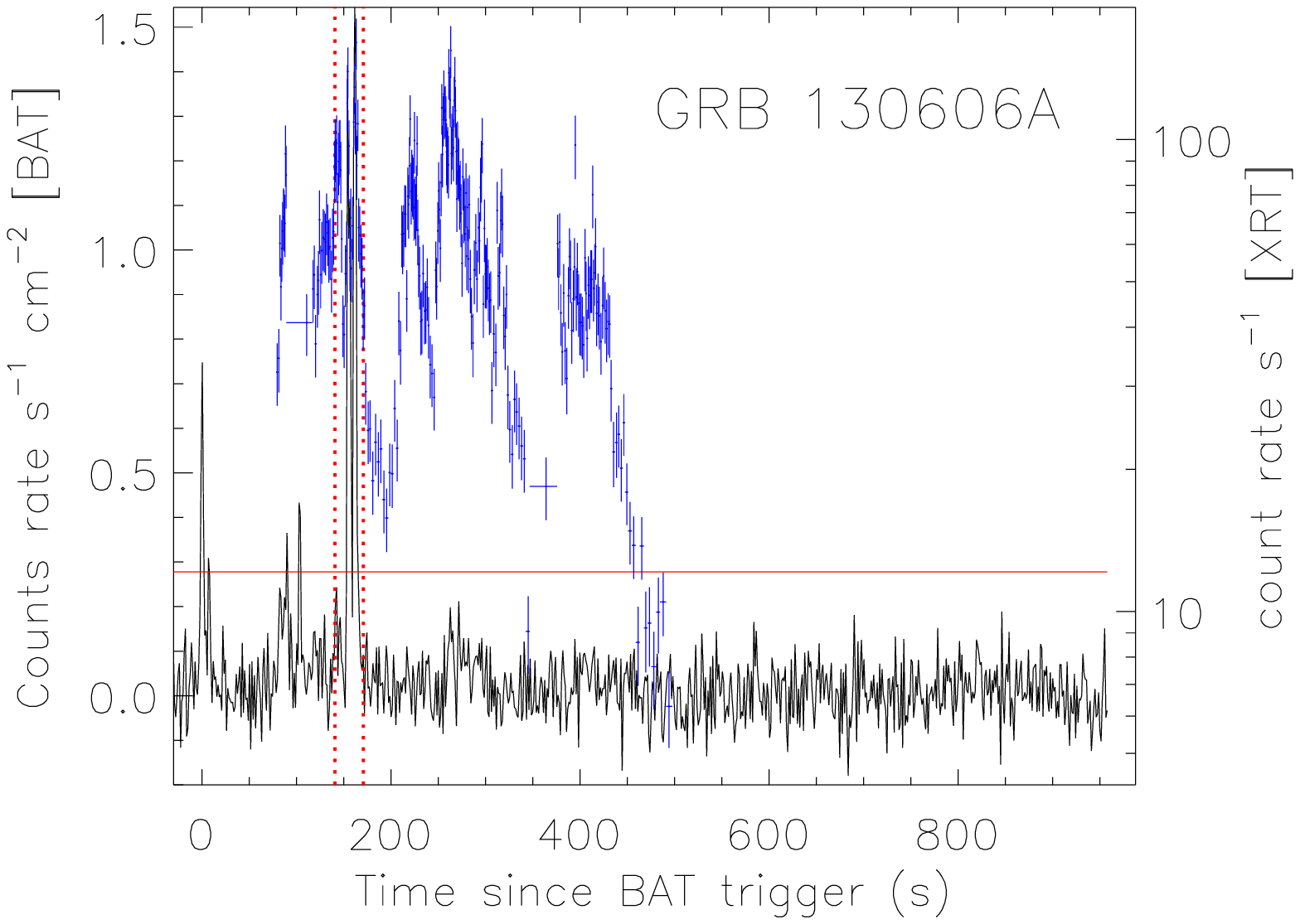}%
\includegraphics[angle=0,scale=0.3]{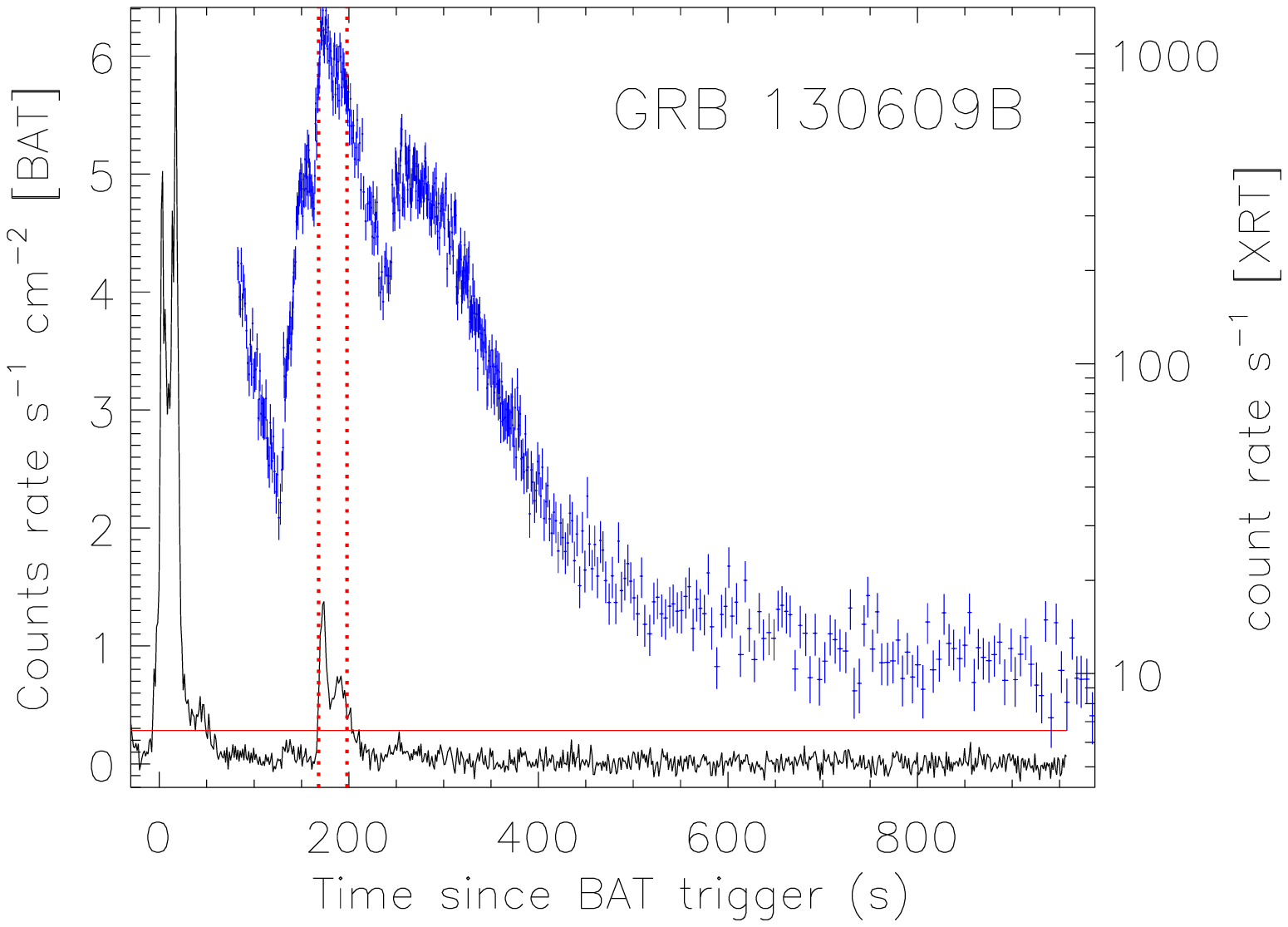}\\

Fig. 1--Continued
\nonumber
\end{figure}

\clearpage

\begin{figure}
\includegraphics[angle=0,scale=0.3]{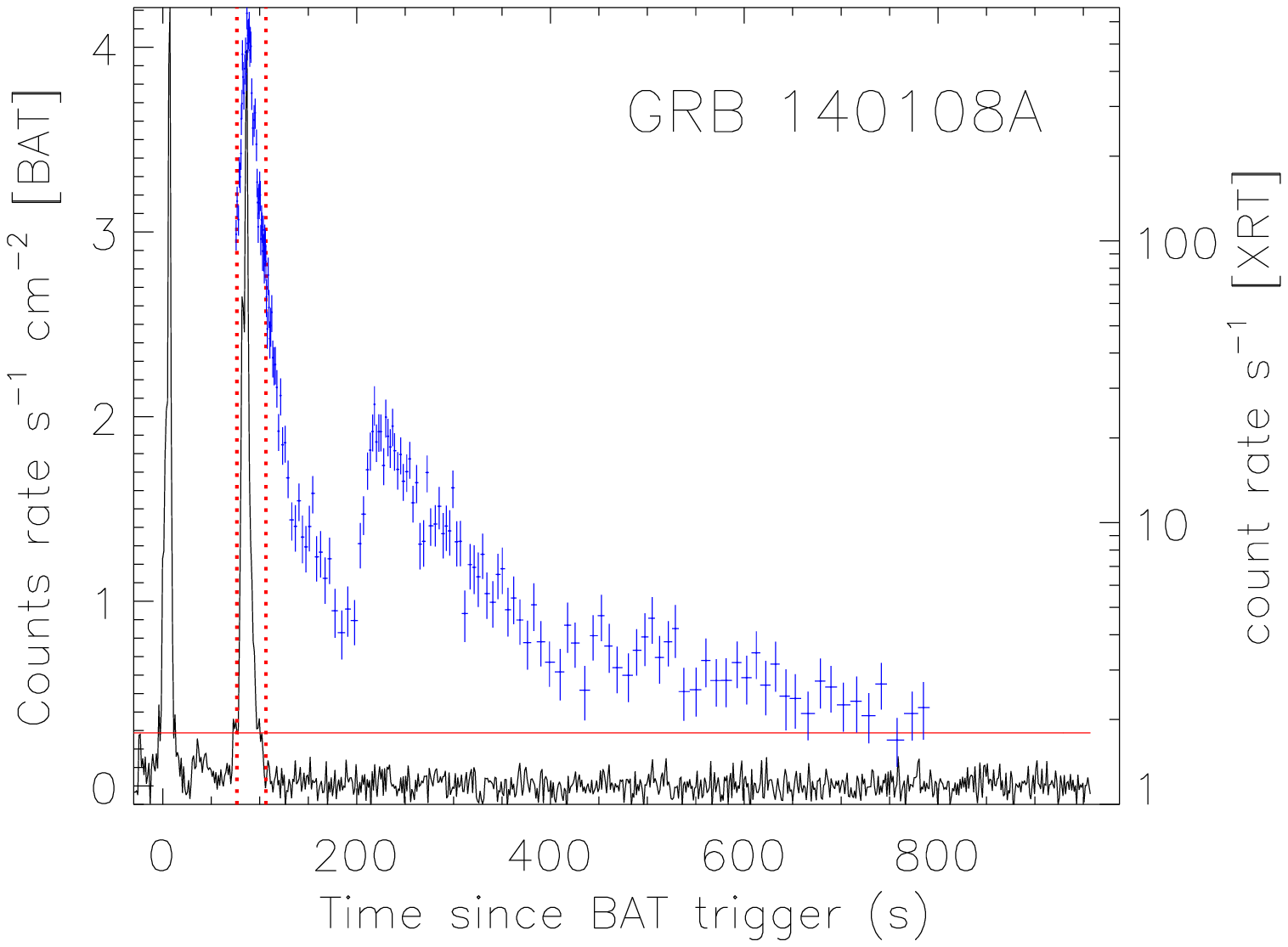}%
\includegraphics[angle=0,scale=0.3]{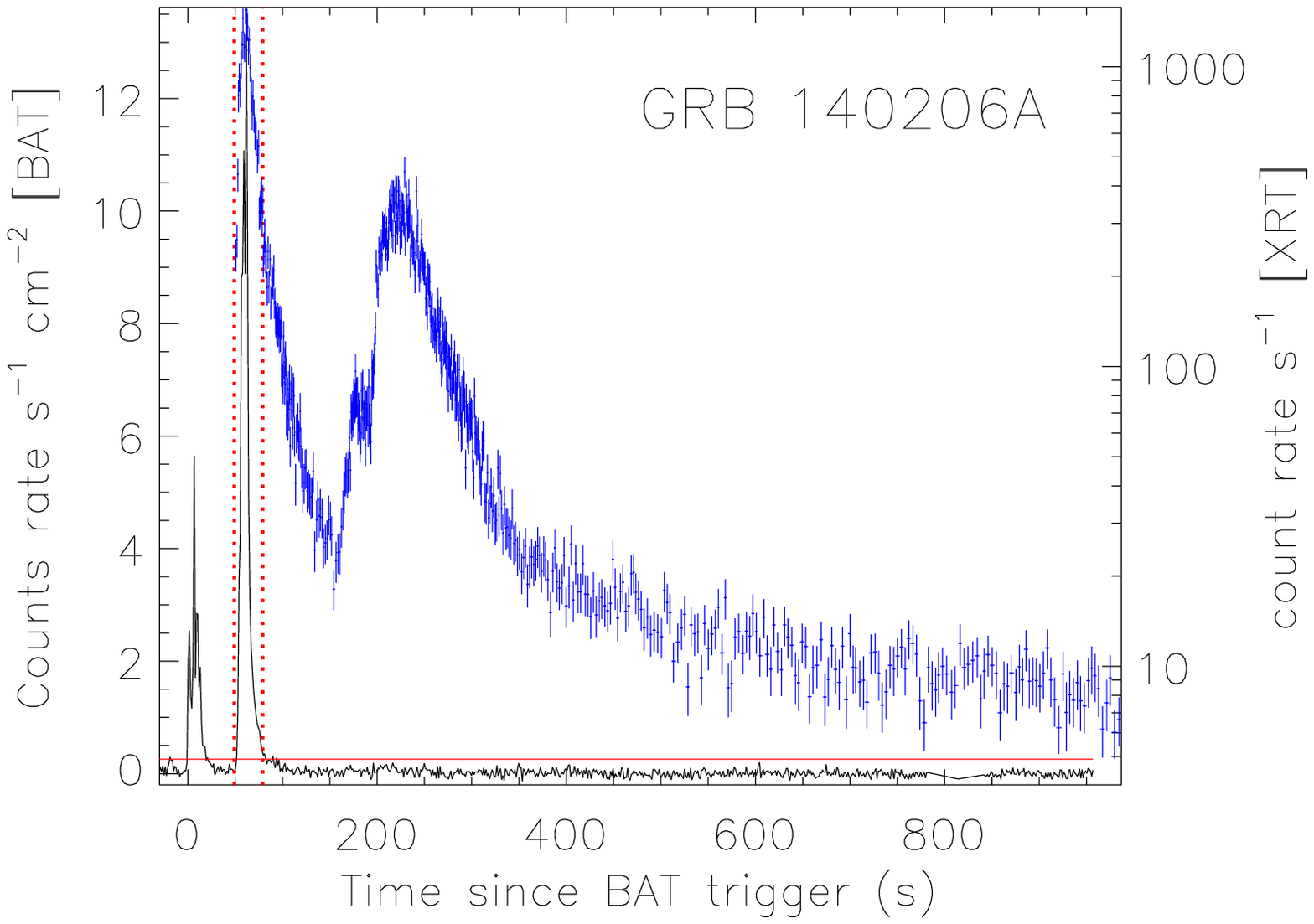}\\
\hfill
Fig. 1--Continued
\nonumber
\end{figure}

\begin{figure}
\includegraphics[angle=270,scale=0.3]{121123A.eps}%
\includegraphics[angle=270,scale=0.3]{121211A.eps}\\
\includegraphics[angle=270,scale=0.3]{140206A.eps}%
\hfill

Fig. 3--Continued
\nonumber
\end{figure}

\begin{figure}
\includegraphics[angle=270,scale=0.3]{100725B.eps}%
\includegraphics[angle=270,scale=0.3]{110801A.eps}\\
\includegraphics[angle=270,scale=0.3]{121217A.eps}%
\includegraphics[angle=270,scale=0.3]{130514A.eps}\\
\includegraphics[angle=270,scale=0.3]{130609B.eps}%
\hfill

Fig. 4--Continued
\nonumber
\end{figure}

\clearpage

\begin{table}
\centering
\caption{Results of our fits  with the single power-law model to the joint spectra observed with BAT and XRT for the 8 flares.}
\begin{tabular}{llccc}
\hline
GRB & interval(s) & $\Gamma$ & $N_{H,22}^{\rm host}$  & $\chi^2$/bins($\rm \chi^2_{red}$)\\
\hline
060607A & 82-112&$1.60_{-0.05}^{+0.05}$&$1.39_{-0.38}^{+0.44}$&176/143(1.26)\\
100212A & 70-90&$1.75_{-0.05}^{+0.06}$&$0.14_{-0.03}^{+0.03}$&117/136(0.88)\\
100728A &107-137&$1.38_{-0.01}^{+0.01}$&$0.36_{-0.04}^{+0.04}$&209/195(1.09)\\
100901A &381-411&$1.34_{-0.03}^{+0.03}$&$0.54_{-0.15}^{+0.19}$&139/136(1.05)\\
111103B &100-130&$1.68_{-0.03}^{+0.03}$&$0.33_{-0.04}^{+0.05}$&143/156(0.94)\\
120514A &135-165&$1.88_{-0.05}^{+0.05}$&$1.47_{-0.15}^{+0.16}$&151/169(0.91)\\
130606A &145-160&$1.03^{+0.03}_{-0.03}$&$3.89^{+3.88}_{-2.70}$&115/95(1.26)\\
140108A &81-96  &$1.40^{+0.02}_{-0.02}$&$0.08_{-0.08}^{+0.20}$&100/111(0.93)\\
\hline
\end{tabular}
\label{Table_8PL}
\end{table}

\begin{table}
\centering
\caption{Results of our fits with the Band function model\tablenotemark{1} to the joint spectra observed with BAT and XRT for the 11 flares.}
\begin{tabular}{llccccc}
\hline
GRB &interval(s) &$\alpha$ &$\beta$ &$E_{\rm p}$(keV) &$N^{\rm host}_{H,22}$&$\chi^2$/bins($\rm \chi^2_{red}$)\\
\hline

090715B   &    61-91      &   $-0.86_{-0.13}^{+0.32}$ &   $-1.81_{-0.12}^{+0.14}$ &    $16.65_{-9.59}^{+9.59}$    &    $2.04_{-0.83}^{+0.80}$ &   211/185(1.17)\\
100619A   &    85-100     &   $-1.00$                 &   $-1.94_{-0.05}^{+0.05}$ &   $12.47_{-1.87}^{+1.86}$ &    $0.55_{-0.09}^{+0.10}$ &    116/116(1.03)\\
100704A   &    160-190    &   $-1.09_{-0.17}^{+0.38}$ &   $-2.64_{-0.13}^{+0.12}$ &   $6.24_{-3.20}^{+3.20}$  &    $0.39_{-0.10}^{+0.07}$ &    203/204(1.03)\\
100814A   &    135-165    &   $-1.00$                 &   $-1.77_{-0.10}^{+0.08}$ &    $13.94_{-3.25}^{+3.35}$    &    $0.23_{-0.10}^{+0.11}$ &   159/162(1.01)\\
100906A   &    100-130    &   $-1.00$                 &   $-2.64_{-0.07}^{+0.07}$ &    $6.00_{-0.54}^{+0.52}$ &    $4.40_{-0.89}^{+1.01}$ &   198/172(1.18)\\
110102A  &  198-213  & $-0.49\pm{0.06}$         &  $-1.33\pm{0.03}$         &  $10.00$                        &  $0.12$                 & 129/123(1.08)\\
110119A   &    190-215    &   $-1.00$                 &   $-1.78_{-0.08}^{+0.06}$ &   $13.41_{-2.10}^{+ 2.44}$     &  $0.18_{-0.03}^{+0.03}$   &  223/180(1.27)\\
111123A   &    135-165    &   $-1.00$                 &   $-1.65_{-0.07}^{+0.06}$ &   $17.12_{-3.42}^{+3.60}$ &    $0.20_{-0.03}^{+0.03}$ &    192/195(1.00)\\
121123A  &  230-260  & $-0.90\pm{0.03}$        &  ..                       &  $ 57.54_{-4.04}^{+4.59}$    &  $0.10_{-0.03}^{+0.04}$ & 173/183(0.97)\\
121211A  &  158-188  & $-1.12^{+0.07}_{-0.04}$  &  $-2.39^{+0.13}_{-0.16}$  &  $10.00$                     &  $1.15_{-0.22}^{+0.26}$ & 172/173(1.03)\\
140206A  &  53-68    & $-0.93^{+0.05}_{-0.04}$  &  ..                       &  $99.32^{+10.56}_{-8.52}$    &  $11.16_{-3.49}^{+4.38}$& 142/135(1.09)\\
\hline
\end{tabular}
\tablenotetext{}{(1) Parameters without error are not well constrained by data. GRB 121123A and GRB 140206A favor cutoff Power-law model, which is a good approximation for Band function.}
\label{Table_11Band}
\end{table}

\begin{table}
\footnotesize
\centering
\caption{Results of our fits with the single power-law plus black body model to the joint spectra observed with BAT and XRT for the 13 flares.}
\begin{tabular}{lllccccccc}
\hline
GRB\tablenotemark{a} &$z^{\rm ref}$&interval(s)&$\Gamma_{\rm PL}$&$K_{\rm PL}$&$kT$(keV)&$K_{\rm BB}$& $N_{\rm H,22}^{\rm host}$& $\chi^2$/bins($\rm \chi^2_{r}$)\\
\hline
060510B & 4.9\tablenotemark{1}  & 290-320&$ 1.50\pm {0.04}$       & $0.66\pm{0.08}$         &$ 1.41_{-0.20}^{+0.32}$    &$0.03^{+0.007}_{-0.006}$&$ 3.37_{-1.31}^{+1.52}$&158/168(0.97)\\
060526  &   3.21\tablenotemark{2}   &   236-266 &   $1.71_{-0.05}^{+0.06}$  &   $1.68_{-0.27}^{+0.34}$  &    $1.56_{-0.22}^{+0.64}$ &    $0.11_{-0.02}^{+0.04}$ &   $5.33_{-1.32}^{+1.72}$  &   145/163(0.92)\\
060814  &   0.84\tablenotemark{3}   &   121-151 &   $1.63_{-0.04}^{+0.04}$  &   $2.02_{-0.26}^{+0.26}$  &    $1.90_{-0.39}^{+0.43}$ &    $0.07_{-0.02}^{+0.03}$ &   $1.28_{-0.21}^{+0.24}$  &   147/146(1.04)\\
061121  &   1.31\tablenotemark{4}   &   65-90   &   $1.32_{-0.02}^{+0.02}$  &   $7.86_{-0.65}^{+0.70}$  &    $2.74_{-0.38}^{+0.31}$ &    $0.53_{-0.18}^{+0.18}$ &   $2.23_{-0.39}^{+0.47}$  &   126/135(0.97)\\
070129  &   ..  &   300-330 &   $1.65_{-0.05}^{+0.05}$  &   $1.16_{-0.16}^{+0.16}$  &   $1.53_{-0.26}^{+0.49}$  &    $0.03_{-0.01}^{+0.01}$ &    $0.25_{-0.04}^{+0.04}$ &   199/193(1.06)\\
080810  &   3.35\tablenotemark{5}   &   90-120  &   $1.48_{-0.05}^{+0.05}$  &   $0.50_{-0.08}^{+0.08}$  &    $1.22_{-0.18}^{+0.30}$ &    $0.03_{-0.01}^{+0.01}$ &   $1.63_{-0.72}^{+0.82}$  &   169/156(1.12)\\
080928  &   1.69\tablenotemark{6}   &   193-223 &   $1.60_{-0.04}^{+0.04}$  &   $2.06_{-0.27}^{+0.29}$  &    $1.38_{-0.15}^{+0.19}$ &    $0.12_{-0.02}^{+0.02}$ &   $1.14_{-0.30}^{+0.35}$  &   162/181(0.92)\\
090709A & ..   &74-104  &$ 1.32\pm{0.03}$        & $1.44_{-0.14}^{+0.16}$  &$ 2.27_{ -0.21}^{+0.18}$   &$0.15\pm{0.03}$         &$ 2.28_{-0.57}^{+0.70}$&196/173(1.17)\\
100725B &   ..  &   200-230 &   $2.40_{-0.08}^{+0.08}$  &   $9.63_{-2,00}^{+2.18}$  &   $1.10_{-0.14}^{+0.24}$  &    $0.13_{-0.03}^{+0.03}$ &    $0.64_{-0.08}^{+0.09}$ &   176/176(1.03)\\
110801A & 1.86\tablenotemark{7}  &344-374 &$ 1.63\pm{0.05}$        & $1.67_{-0.27}^{+0.31}$  &$ 1.24_{-0.11}^{+ 0.15}$   &$0.11\pm{0.01}$         &$ 1.51_{-0.47}^{+0.62}$&226/203(1.15)\\
121217A & ..   &715-745 &$1.44_{-0.03}^{+0.04}$  & $1.43_{-0.13}^{+0.25}$  &$2.61_{-0.21}^{+0.20}$     &$0.14_{-0.03}^{+0.02}$  & $0.01_{-0.01}^{+0.12}$&193/162(1.23)\\
130514A & 3.6\tablenotemark{8}  &95-125  &$ 1.67\pm{0.03}$        & $2.48_{-0.24}^{+0.26}$  &$ 2.47_{-0.24}^{+0.21}$    &$0.16\pm{0.03}$         &$0.09\pm{0.05}$        &205/199(1.06)\\
130609B & ..   &178-208 &$ 1.79\pm{0.03}$        & $7.29_{-0.69}^{+0.75}$  &$ 2.07_{-0.16}^{+0.14}$    &$0.34\pm{0.06}$         &$ 0.16\pm{0.04}$       &215/195(1.13)\\
\hline
\end{tabular}
\tablenotetext{}{(1)Price et al.(2006); (2)Berger et al.(2006); (3)Thoene et al.(2007); (4)Bloom et al.(2006); (5)Prochaska et al.(2008); (6)Vreeswijk et al.(2008); (7)Cabrera Lavers et al.(2011);(8) Schmidl et al.(2013).}
\label{Table_13BB}
\end{table}

\begin{table}
\centering
\caption{Derived fireball properties of the 8 flares that have detection of a thermal component and redshift measurement in our sample.}
\begin{tabular}{lccccccccc}
\hline
GRB&$L_{\rm BB,50}$& $E_{\rm p}{'}$& $\Gamma_{\rm ph}$ & $R_{\rm ph,13}$& $R_{\rm sat,10}$ & $R_{0,7}$\tablenotemark{a} \\
\hline
060510B &$  6.23    \pm 1.45 $&$    23.46   \pm 5.32    $&  141.04  &   1.42    &   1.25    &   8.85  \\
060526  &$  8.36    \pm 3.35 $&$    18.52   \pm 7.60   $&  104.9   &   1.96    &   6.26    &   59.69  \\
060814  &$  0.19    \pm 0.07 $&$    9.86    \pm 2.23    $&  57.97   &   0.57    &   0.56    &   9.61  \\
061121  &$  4.55    \pm 1.54 $&$    17.88   \pm 2.02    $&  124.83  &   1.85    &   1.18    &   9.48  \\
080810  &$  2.05    \pm 0.42 $&$    14.97   \pm 3.68    $&  98.01   &   1.39    &   1.22    &   12.48 \\
080928  &$  1.92    \pm 0.29 $&$    10.47   \pm 1.44    $&  71.77   &   2.01    &   3.74    &   52.16 \\
110801A &$  2.22    \pm 0.20 $&$    10.00   \pm 1.21    $&  68.34   &   2.26    &   5.50    &   80.46 \\
130514A &$  16.01   \pm3.00 $&$     32.04   \pm 2.72    $&  153.5   &   1.33    &   3.63    &   23.67 \\
\hline
\end{tabular}
\tablenotetext{a}{Derived from equation (14).}
\label{Table_Fireball}
\end{table}
\end{document}